\documentclass[11pt]{gmarticle}
\usepackage{graphicx,epsfig,epsf,cite}
\usepackage{fancyheadings}
\usepackage{amsmath}
\usepackage{amssymb}
\usepackage[footnotesize]{caption2}
\makeatother 
\begin{document}
\rhead[]{Proceedings Issue: Behavior of Granular Media (2006)}      
\lhead[]{}
\title{\large\bf
  Efficient numerical simulation of granular matter using\\ the Bottom-To-Top
  Reconstruction method
}

\author{
  \normalsize Thomas Schwager and Thorsten P\"oschel\\
  \normalsize Centrum f\"ur Muskuloskeletale Chirurgie, Charit\'e, Campus
  Virchow-Klinikum,\\ 
  \normalsize Augustenburger Platz 1, 13353 Berlin, Germany
}

\date{\empty}

\twocolumn[{\csname @twocolumnfalse\endcsname
  
\maketitle
\begin{abstract}
  The numerical simulation of granular systems of even moderate size is a
  challenging computational problem. In most investigations, either Molecular
  Dynamics or Event-driven Molecular Dynamics is applied. Here we show that in
  certain cases, mainly (but not exclusively) for static granular packings,
  the Bottom-to-top Reconstruction method allows for the efficient simulation
  of very large systems. We apply the method to heap formation, granular flow
  in a rotating cylinder and to structure formation in nano-powders. We also
  present an efficient implementation of the algorithm in C++, including a 
  benchmark.
\end{abstract}
\vspace{1cm}
}]

\section{Introduction}

The intention of numerical particle simulations of granular matter is (at
least) two-fold. First, simulations are frequently helpful to explain
experimental results, that is, to improve our understanding of the static and
dynamic behavior of granular matter and to improve its corresponding
theoretical description. Second, yet more important, particle simulations may
be helpful for engineers to construct or optimize technical devices to handle
and process granular materials. Having a
reliable numerical method at hand, the prediction of granular matter behavior
is even possible if there is no theoretical description due to our
limited understanding or for more fundamental reasons \cite{Goldhirsch:1999}.

The main problem in applying particle simulations as a standard tool in
engineering is the enormous consumption of computational power (see
\cite{algo} for an overview). The most
universal method for particle simulation is {\em Molecular Dynamics} (MD), also
called Discrete Elements Method, where Newton's equation of motion is solved
simultaneously for all particles of the system. The knowledge of the
interaction forces is sufficient, in principle, to simulate any granular 
assembly with arbitrary boundary conditions.  Unfortunately, due to the large
stiffness of the particle, expressed by the Young modulus of the material, MD
simulations of granular matter require a very small integration time step
which implies high computational costs. At present, for systematic
investigations of systems over a longer period, the number of particles is
restricted to significiantly below $10^5$. On the other hand, even a
relatively small container as used for experimental investigations 
comprises much more particles, sometimes billions.

The simulation may be accelerated considerably by assuming that each
contact of particles is an instantaneous event, that is, the duration of
contacts vanishes and in the entire system at any time only two particles
are in contact. Under this condition, {\em Event-driven MD} (EMD) may be
applied. Here, instead of integrating Newton's equation, the evolution of the
system is determined by the
sequence of particle collisions,
where the postcollision velocities are computed as functions of the
precollision velocities and the coefficients of restitution which are
functions of the impact velocity themselves, see \cite{algo}. In the limit of
very dilute systems of stiff particles, called granular gases, both algorithms, MD and EMD,
deliver identical results.  

Using desktop computers, at present we can simulate about $10^7$ particles
using EMD, however, the abovementioned condition of EMD does obviously not allow for its application
as a general simulation tool. EMD, in principle, is not suited to simulate
systems where particles contact each other over a non-vanishing period of time.

A complementary approach is the {\em Bottom-To-Top Reconstruction} method
(BTR). Here the simultaneous numerical integration of the $N$-particle system
is approximated by the sequential numerical integration of a one-particle system
with complex boundary conditions. Similar as EMD, BTR is also much more
efficient than MD at the price of a restricted applicability, see
Sec. \ref{sec:bench}. While the main precondition of EMD is vanishing
contact time, for BTR we have to require that a particle having arrived in a
stable position does not leave this position anymore. That is, complementary
to EMD, BTR requires persistant contacts. There are cases when the application
of BTR leads to unphysical results, however, if the algorithm is applicable,
it leads to a vast increase of numerical efficiency; at present we can simulate
systems of many millions particles on a desktop computer. The range of its
applicability and benchmarks of the algorithm are discussed at the end of this article.

\section{Bottom-to-top Reconstruction}\label{sec:btr}

The idea of BTR goes back to Visscher and
Bolsterli~\cite{VisscherBolsterli:1972} who suggested an
algorithm that allows for the fast simulation of large static
granular packings. 
The fundamental idea of their method is to consider the motion of the
particles sequentially, unlike in MD simulations, where the coupled system of
Newton's equations of motion is solved for all particles simultaneously. 

The deposition of the particles of a granular packing, e.g., a heap on 
the plane $(x,y,0)$, proceeds as follows: the first particle is inserted 
at position $(x_1^{\rm init},y_1^{\rm init},z_1^{\rm init})$. The 
coordinate $z_1^{\rm init}$ should be larger than the expected final height 
of the heap, the coordinates $x_1^{\rm init}$ and $y_1^{\rm init}$ can be 
chosen at random (particles are scattered over a certain area) or can be 
fixed, e.g., $(x_1^{\rm init}, y_1^{\rm init}) = (0,0)$, to simulate the 
build-up of a heap from a point source. The particle then falls until it 
touches the ground at $(x_1^{\rm init}, y_1^{\rm init}, R_1)$. At this position 
the particle remains fixed. Then the second particle is inserted at position 
$(x_2^{\rm init},y_2^{\rm init},z_2^{\rm init})$. It falls down until it 
touches either the ground at $(x_2^{\rm init},y_2^{\rm init},R_2)$ or the 
first particle, whatever happens first. If it touches the ground it remains 
fixed there just like the first particle. If it, however, touches the first 
particle, it rolls down its surface in the downslope direction until it 
either touches the ground (if $R_2\ge R_1$) where it remains fixed or 
(if $R_2<R_1$) until it loses contact at $z_2=R_1$ when both centers are at the same height. From there it falls to the ground where it 
remains fixed. The next particles are treated likewise. A particle 
remains fixed if it either touches the ground or if it attains a local 
minimum where it is supported by two (in two dimensions) or three (in 
three dimensions) other already fixed particles.

Thus, each particle moves according to a set of rules from one state to 
the next. In this sense the algorithm belongs to the class of event-driven 
algorithms.

\begin{figure}[h!]
\newcommand{\TSTPnegdist}{-0.51cm}
\newcommand{\TSTPfigwidth}{1.75cm}

\noindent
\hspace*{-0.2cm}
\begin{tabular}{lllll}
\includegraphics[width=\TSTPfigwidth]{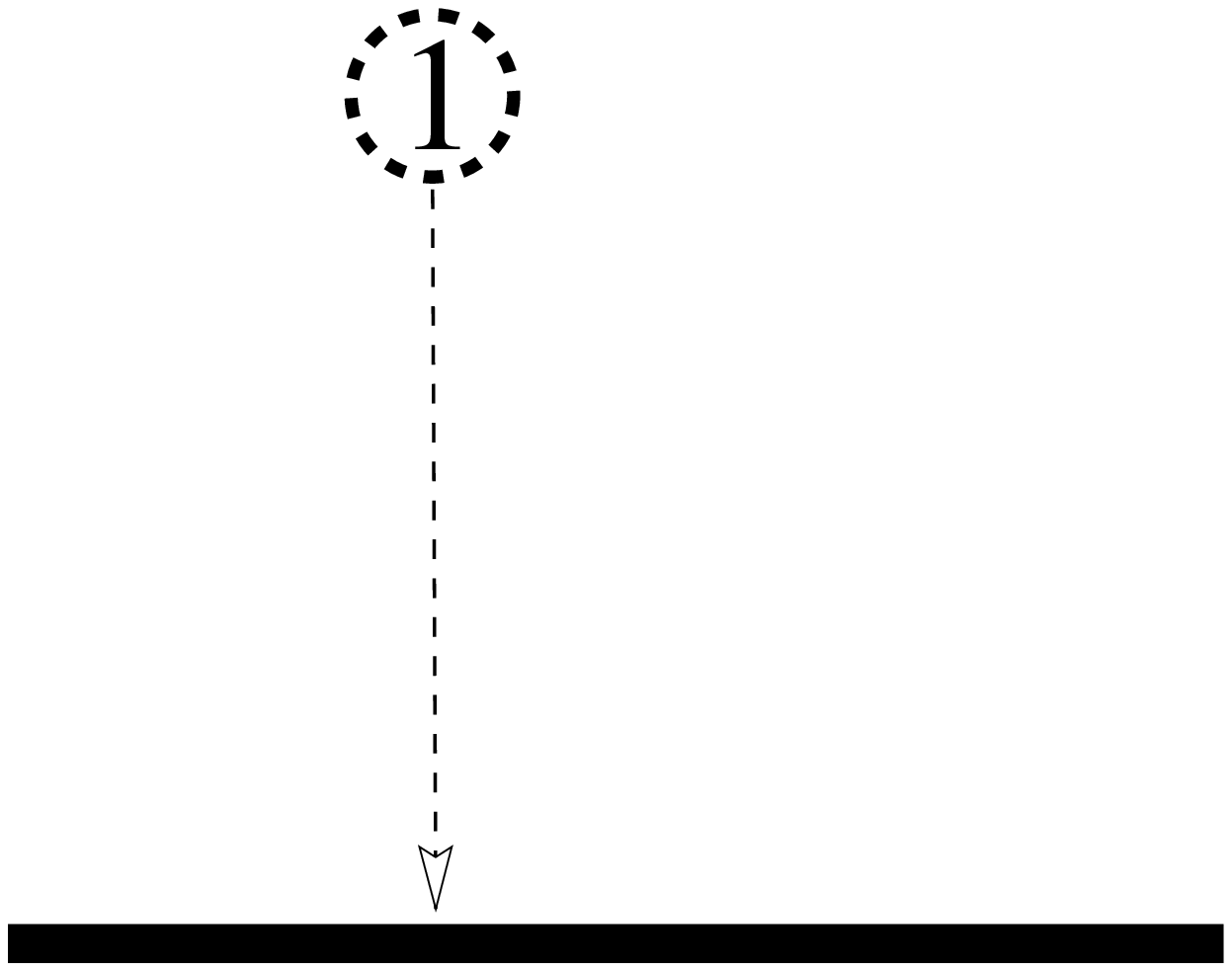}&\hspace*{\TSTPnegdist}
\includegraphics[width=\TSTPfigwidth]{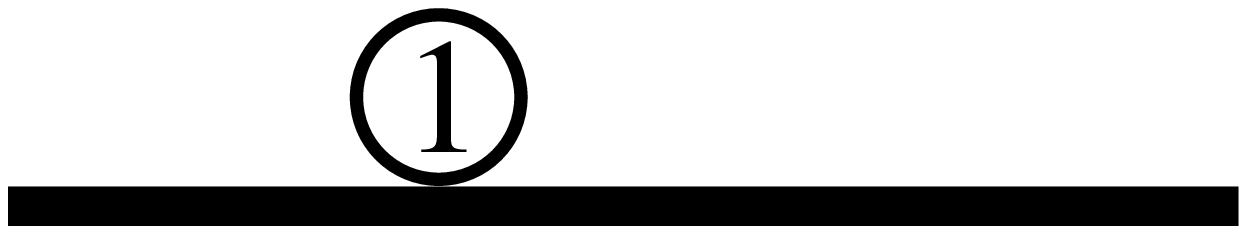}
\\
\multicolumn{5}{p{\columnwidth}}{\footnotesize Particle 1 is inserted and moves downward until it touches the ground.}
\\[0.1cm]
\includegraphics[width=\TSTPfigwidth]{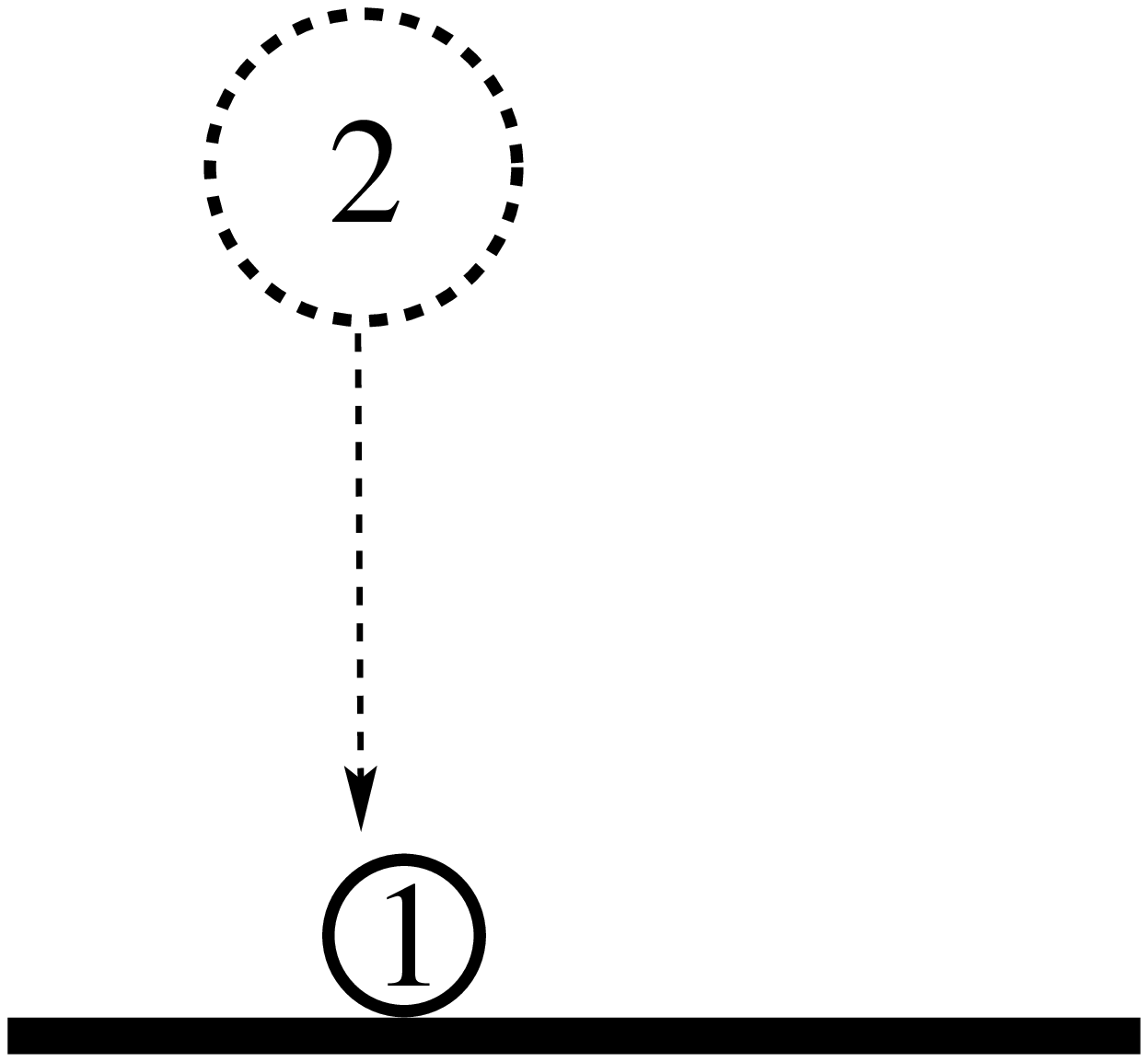}&\hspace*{\TSTPnegdist}
\includegraphics[width=\TSTPfigwidth]{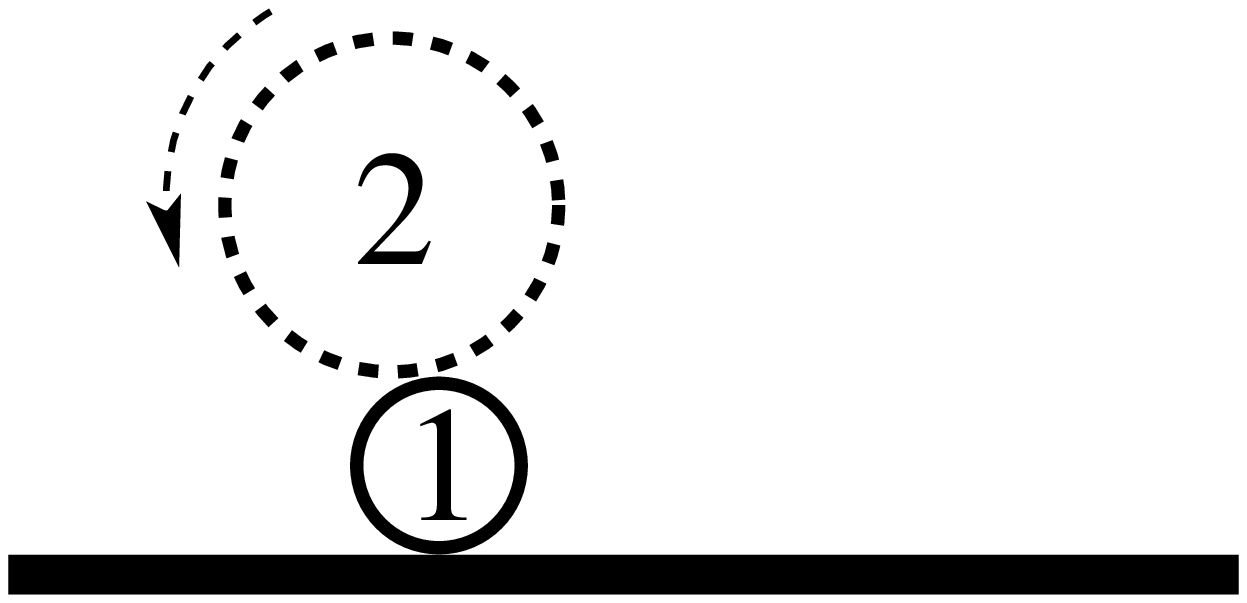}&\hspace*{\TSTPnegdist}
\includegraphics[width=\TSTPfigwidth]{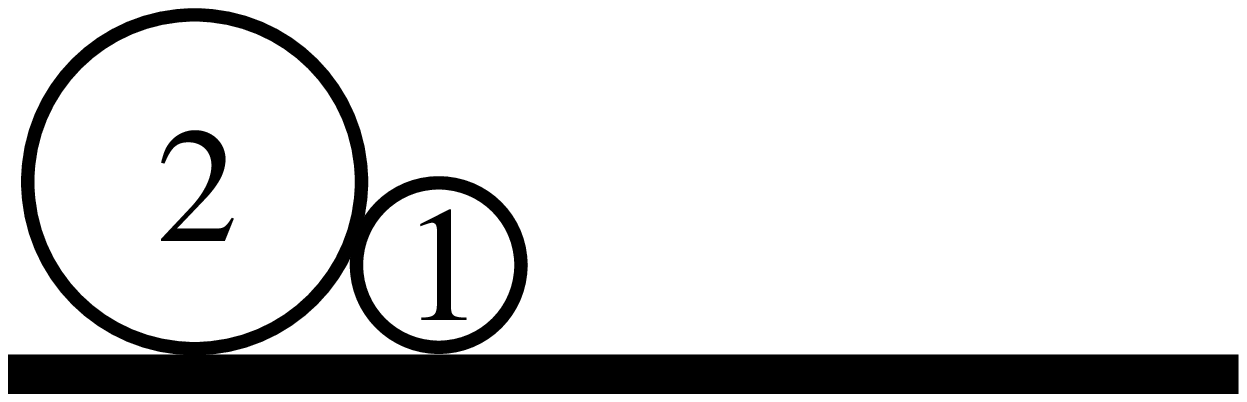}
\\
\multicolumn{5}{p{\columnwidth}}{\footnotesize Particle 2 is inserted and moves downward until it touches particle 1. Then it rolls on the surface of particle 1 until it touches the ground.}
\\[0.6cm]
\includegraphics[width=\TSTPfigwidth]{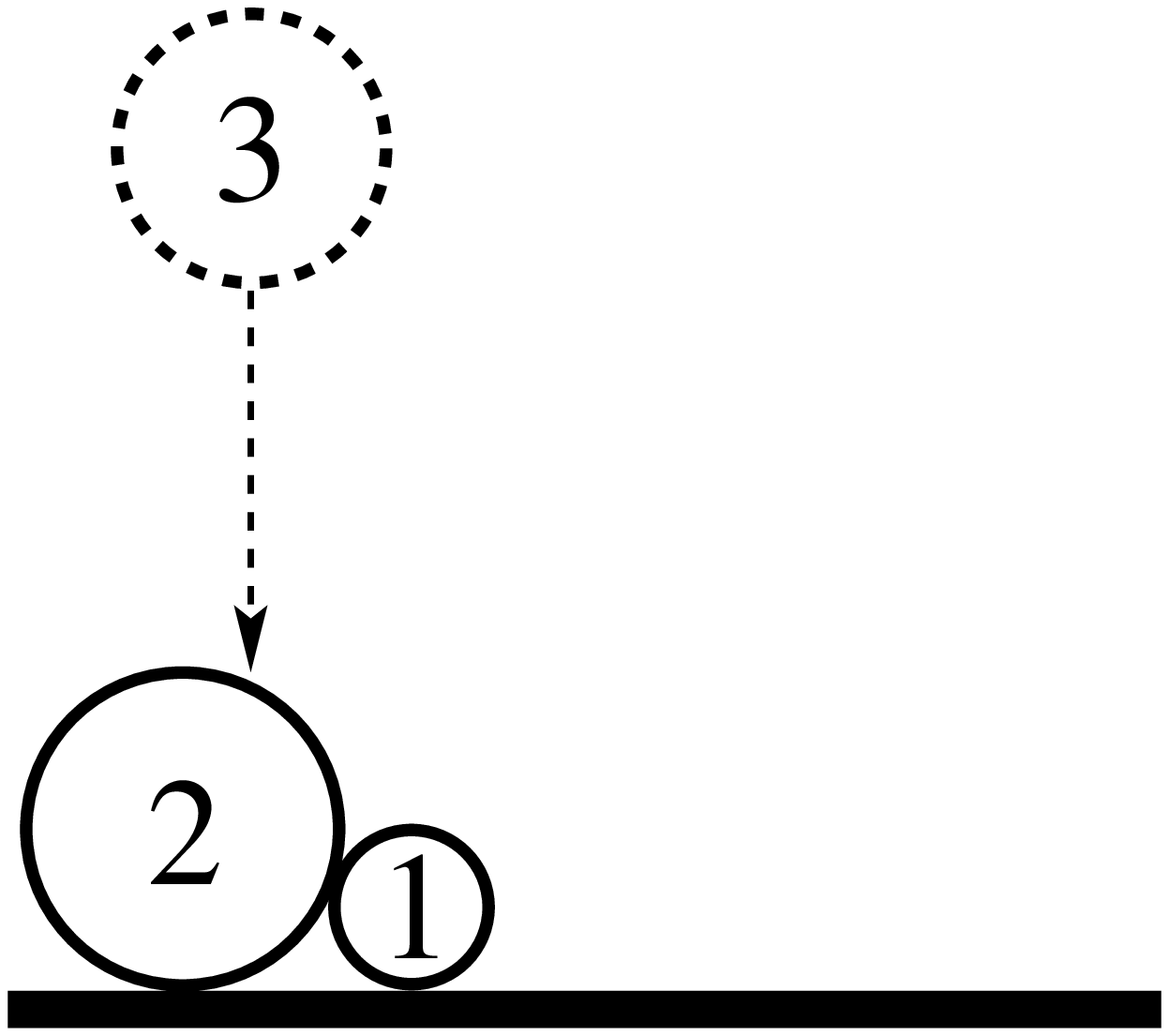}&\hspace*{\TSTPnegdist}
\includegraphics[width=\TSTPfigwidth]{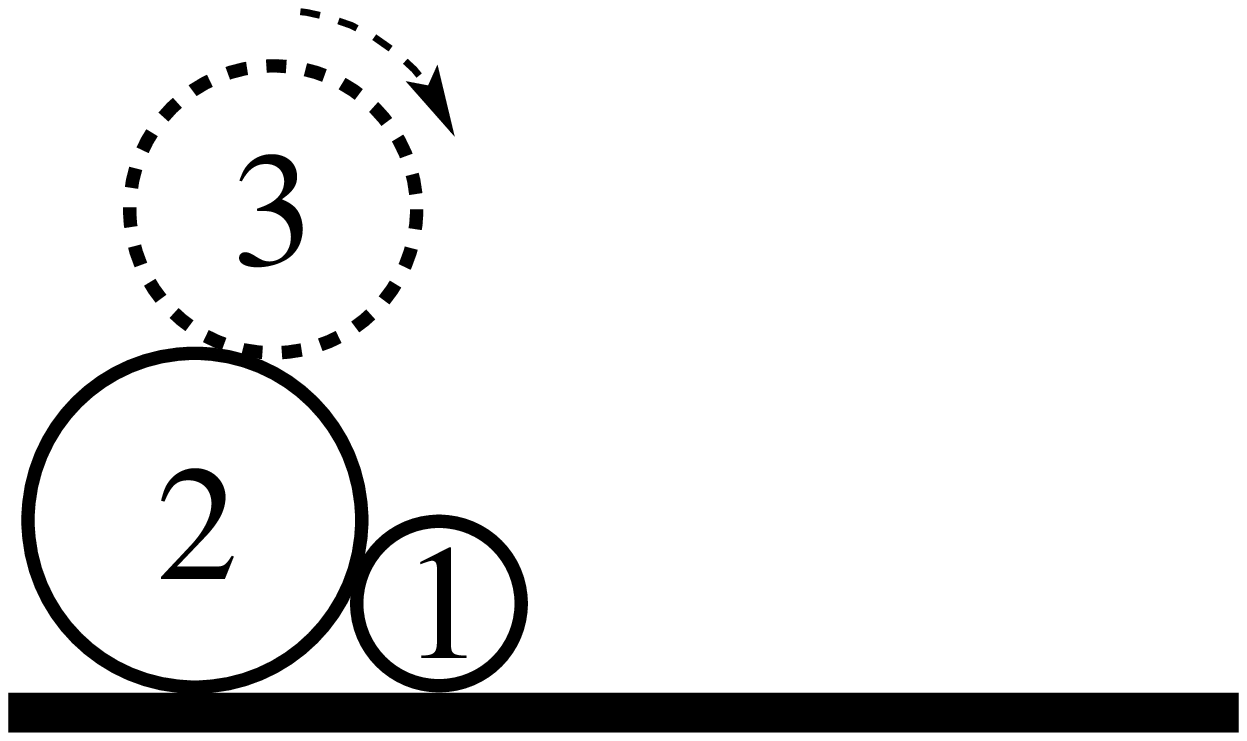}&\hspace*{\TSTPnegdist}
\includegraphics[width=\TSTPfigwidth]{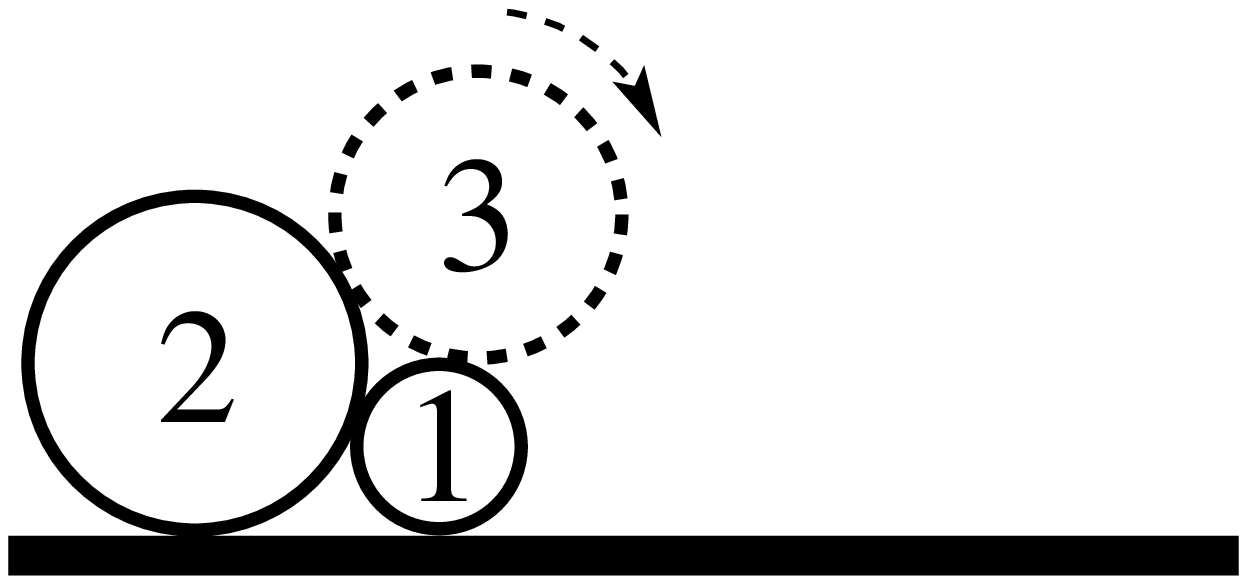}&\hspace*{\TSTPnegdist}
\includegraphics[width=\TSTPfigwidth]{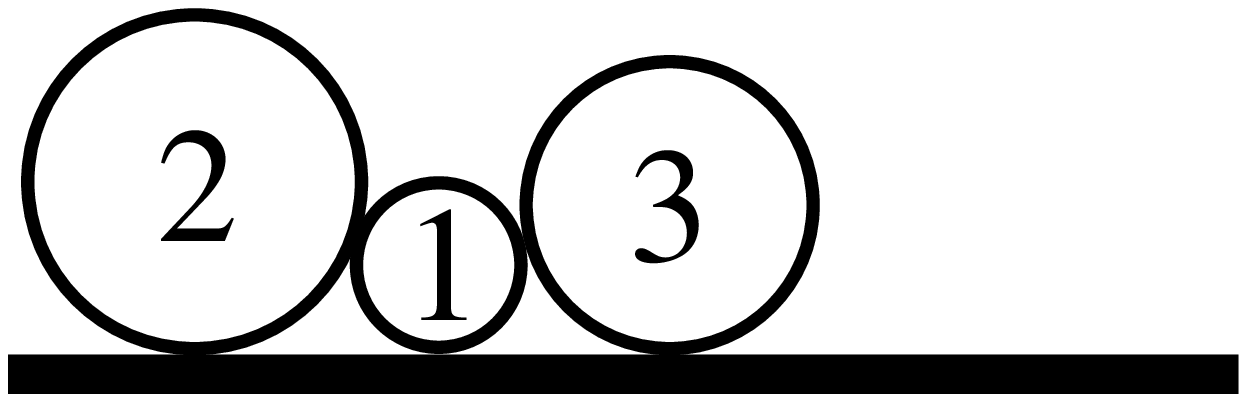}
\\
\multicolumn{5}{p{\columnwidth}}{\footnotesize Particle 3 is inserted and moves downward until it touches particle 2. It then rolls to the right until it touches particle 1. This position is unstable (not a local minimum), thus, the particle continues to roll on particle 1 until it touches the ground.}
\\[0.8cm]
\includegraphics[width=\TSTPfigwidth]{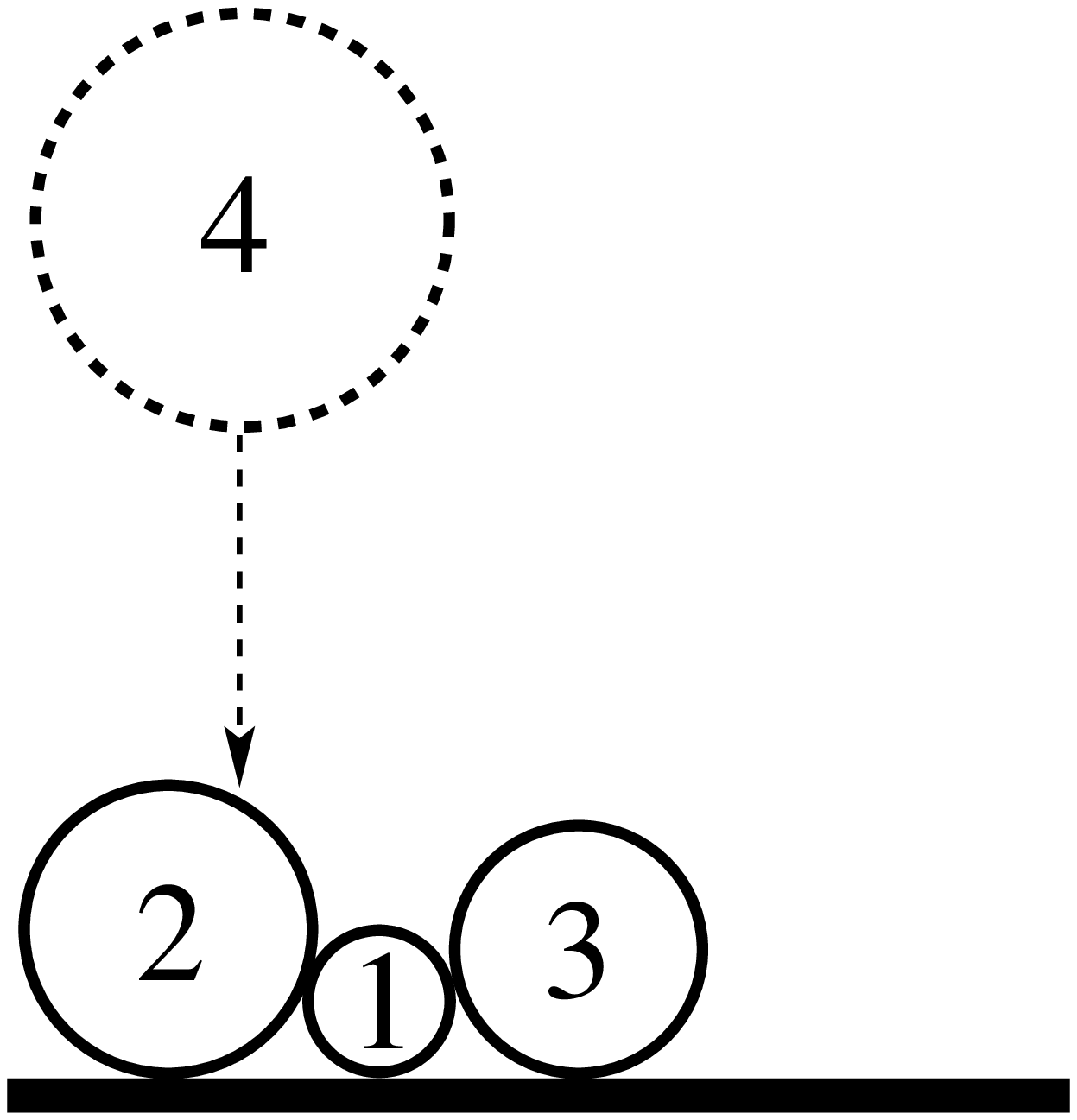}&\hspace*{\TSTPnegdist}
\includegraphics[width=\TSTPfigwidth]{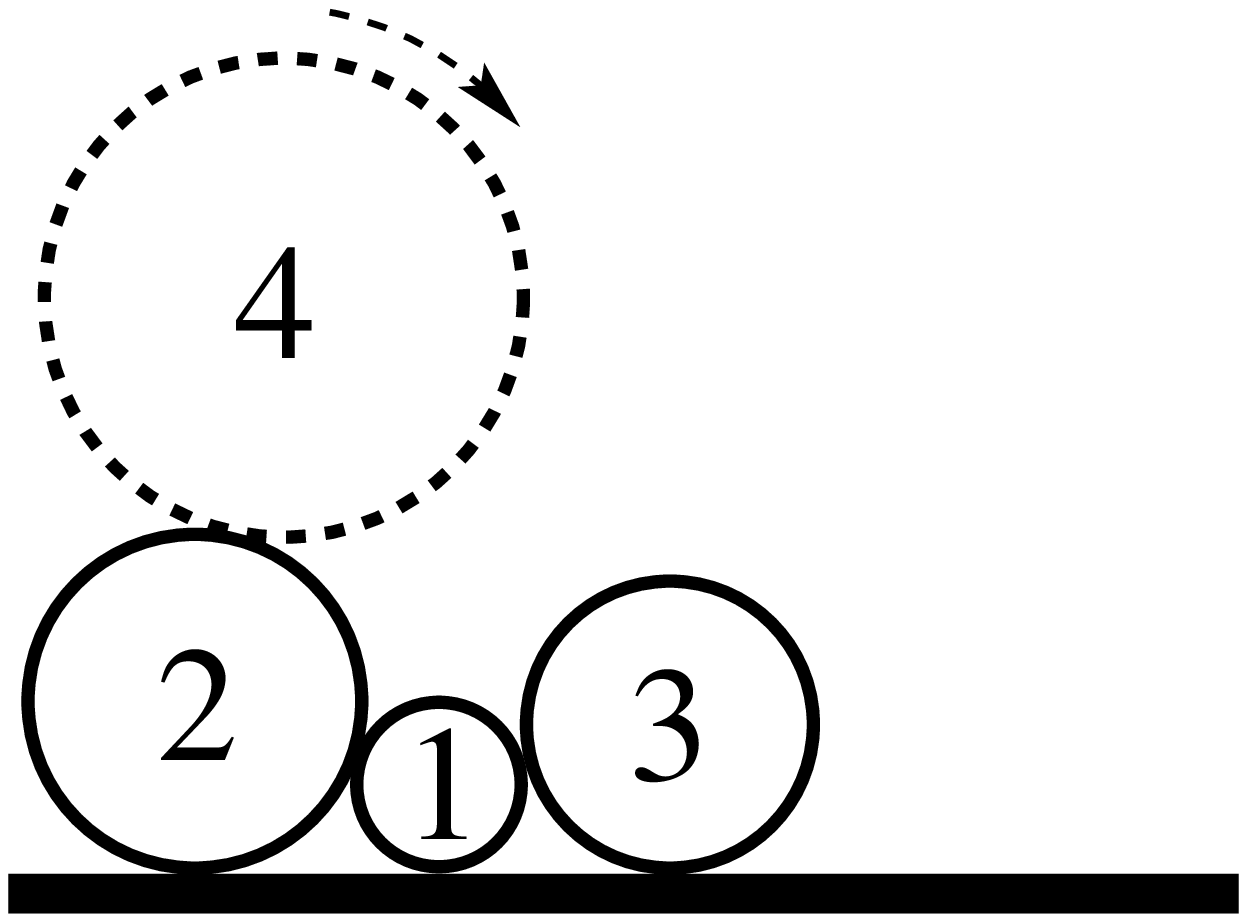}&\hspace*{\TSTPnegdist}
\includegraphics[width=\TSTPfigwidth]{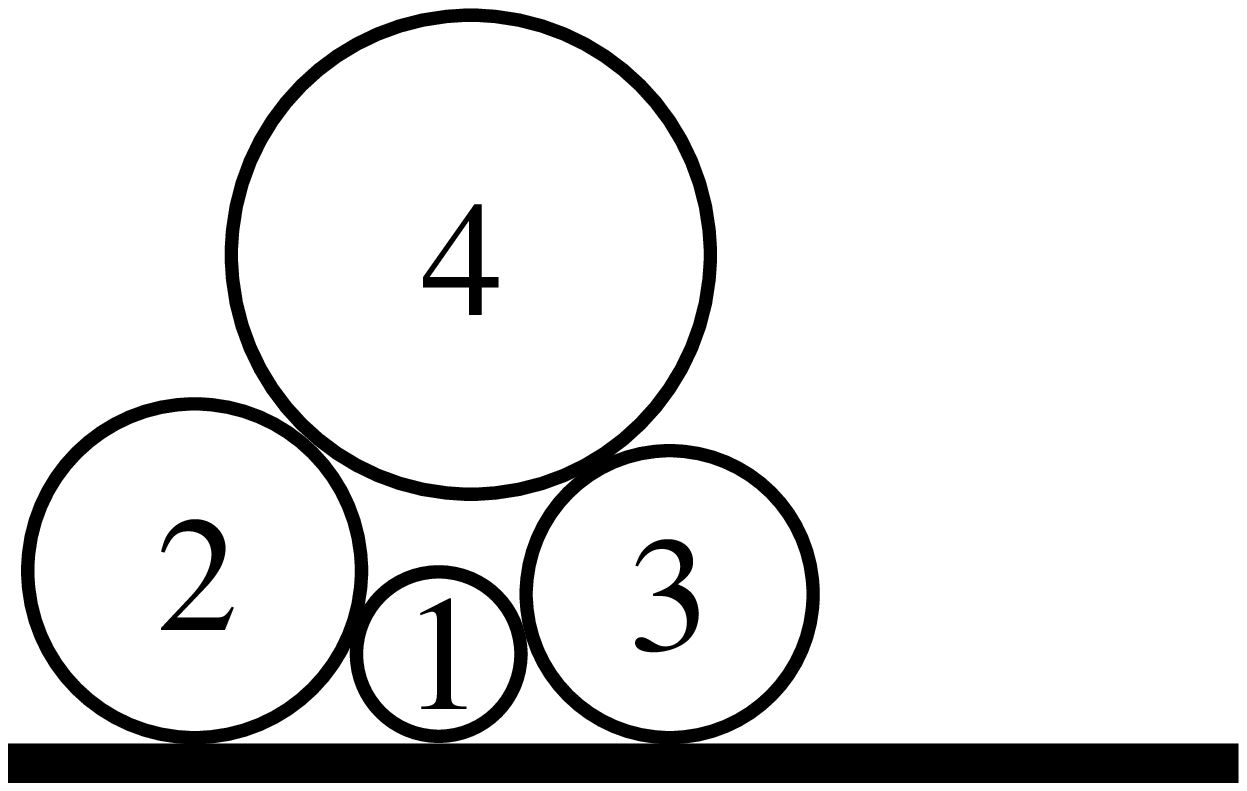}
\\
\multicolumn{5}{p{\columnwidth}}{\footnotesize Particle 4 first touches particle 2 and starts to roll to the right until it touches particle 3. This position is stable (local minimum), thus, particle 4 is fixed at this position.}
\\[0.9cm]
\includegraphics[width=\TSTPfigwidth]{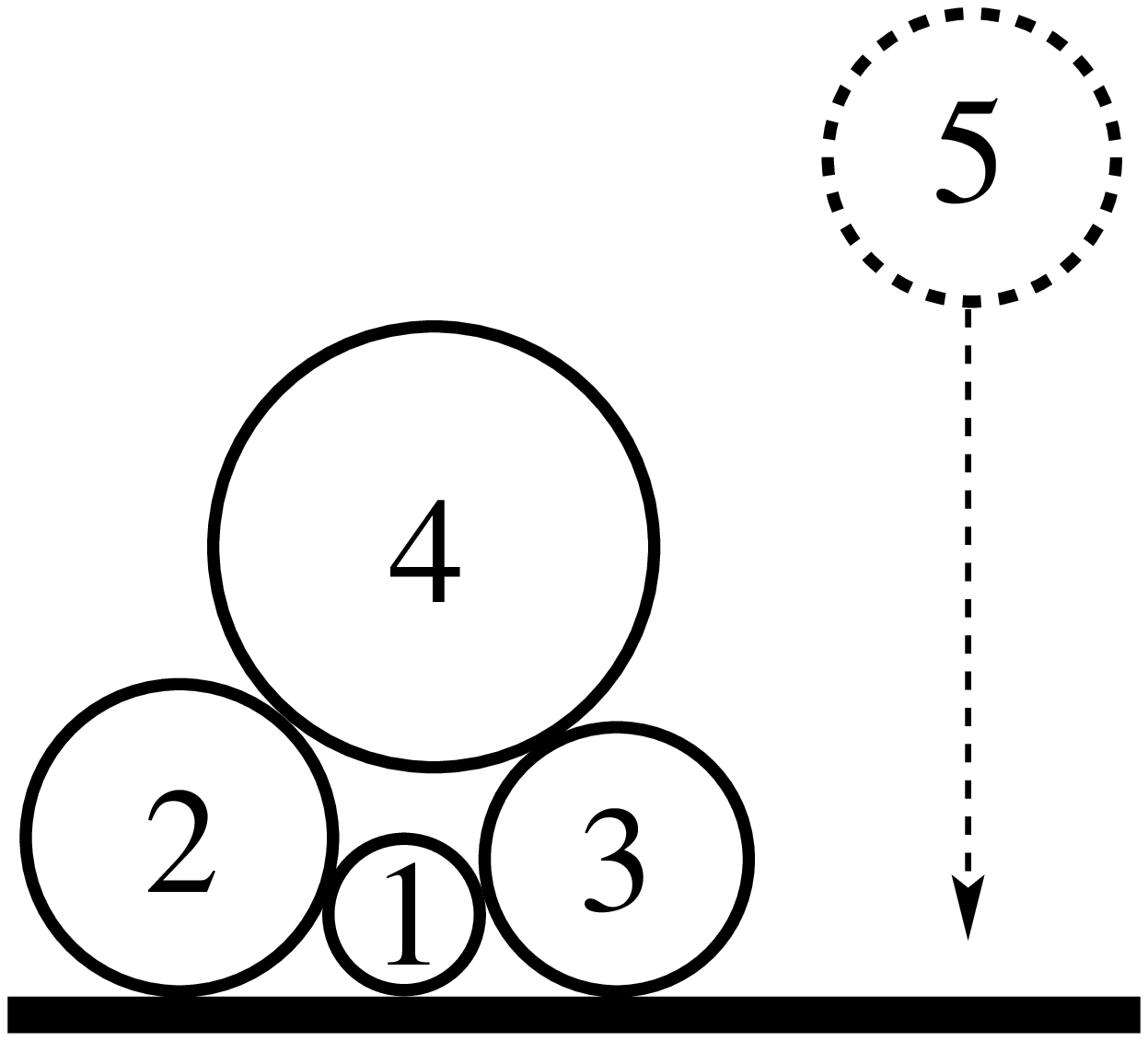}&\hspace*{\TSTPnegdist}
\includegraphics[width=\TSTPfigwidth]{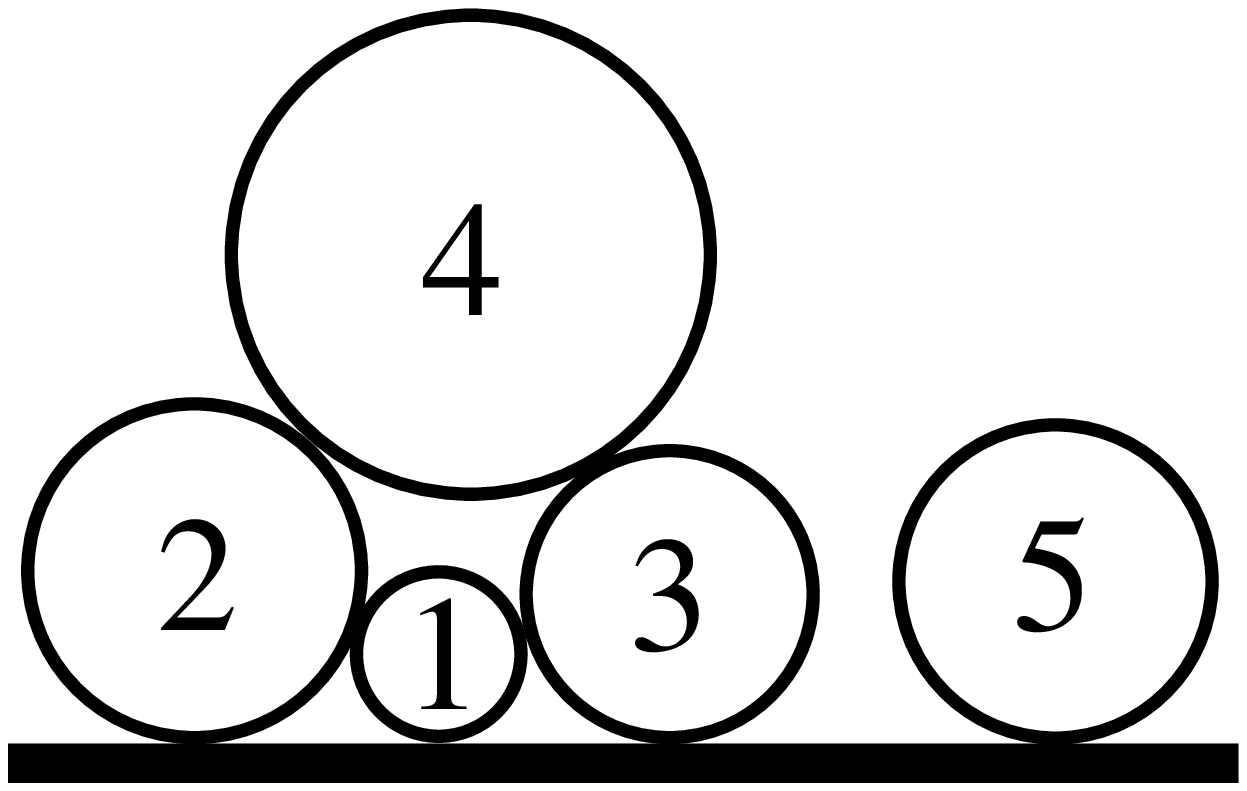}
\\
\multicolumn{5}{p{\columnwidth}}{\footnotesize Particle 5 moves analogously to particle 1 and is fixed on the ground.}
\\[0.2cm]
\includegraphics[width=\TSTPfigwidth]{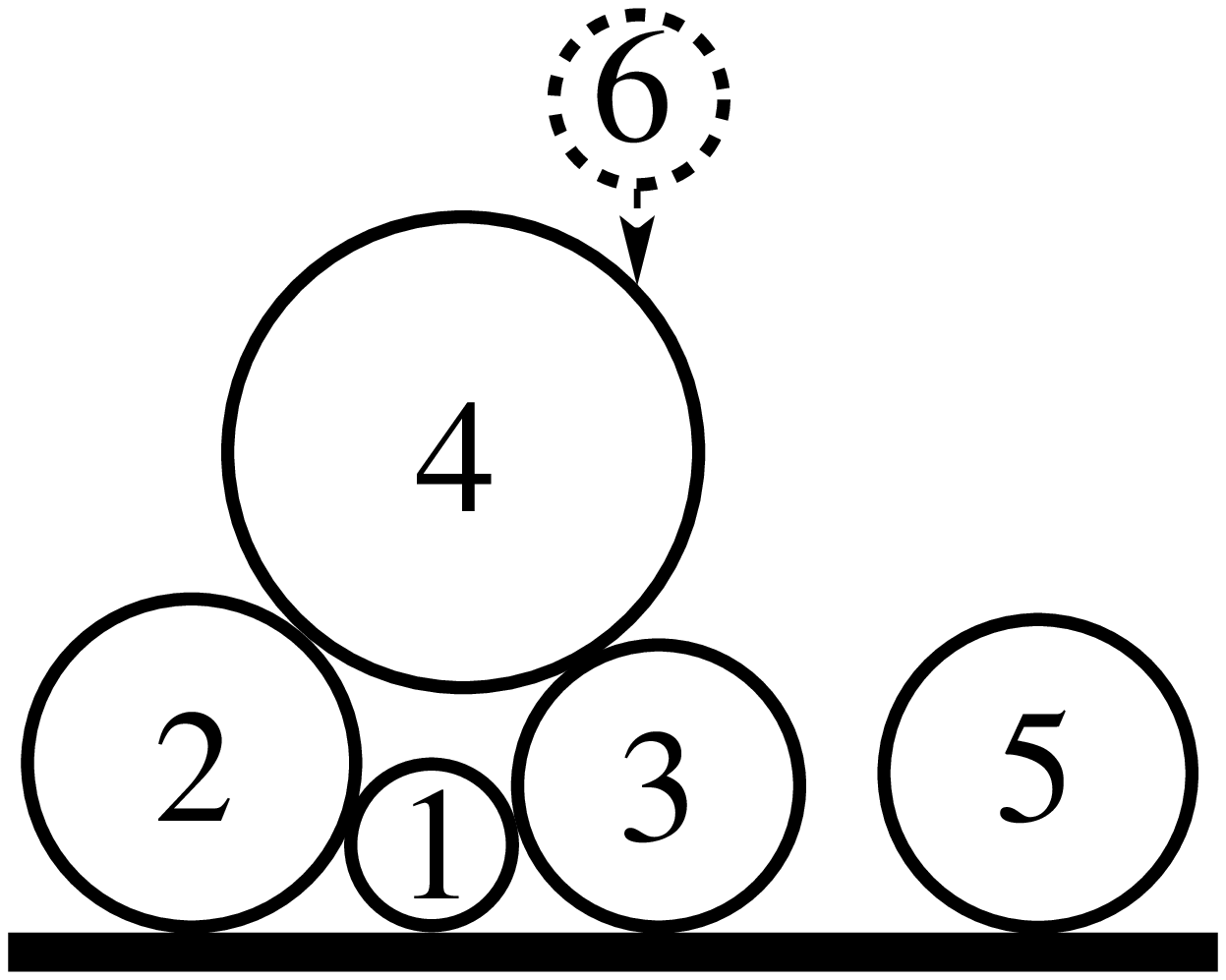}&\hspace*{\TSTPnegdist}
\includegraphics[width=\TSTPfigwidth]{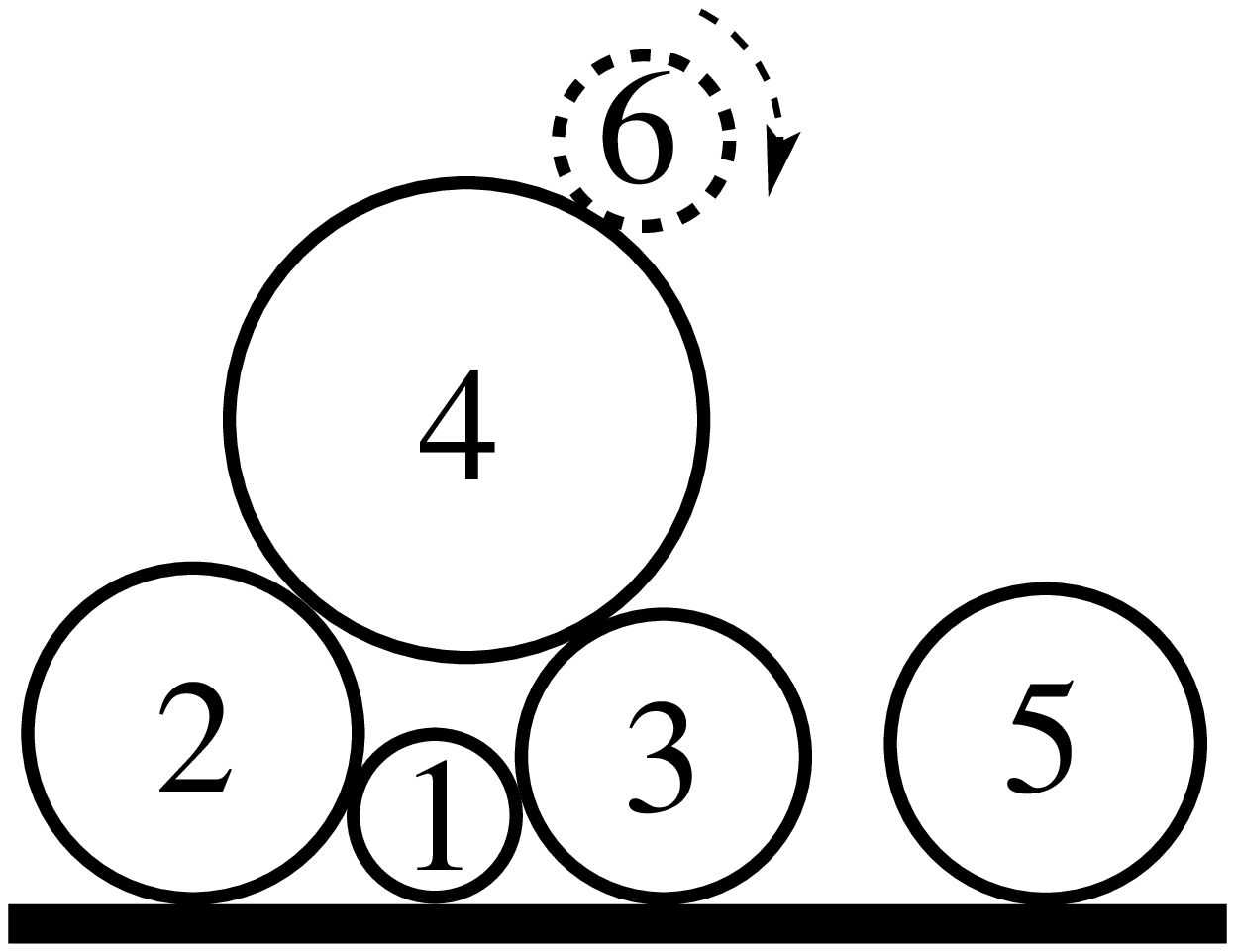}&\hspace*{\TSTPnegdist}
\includegraphics[width=\TSTPfigwidth]{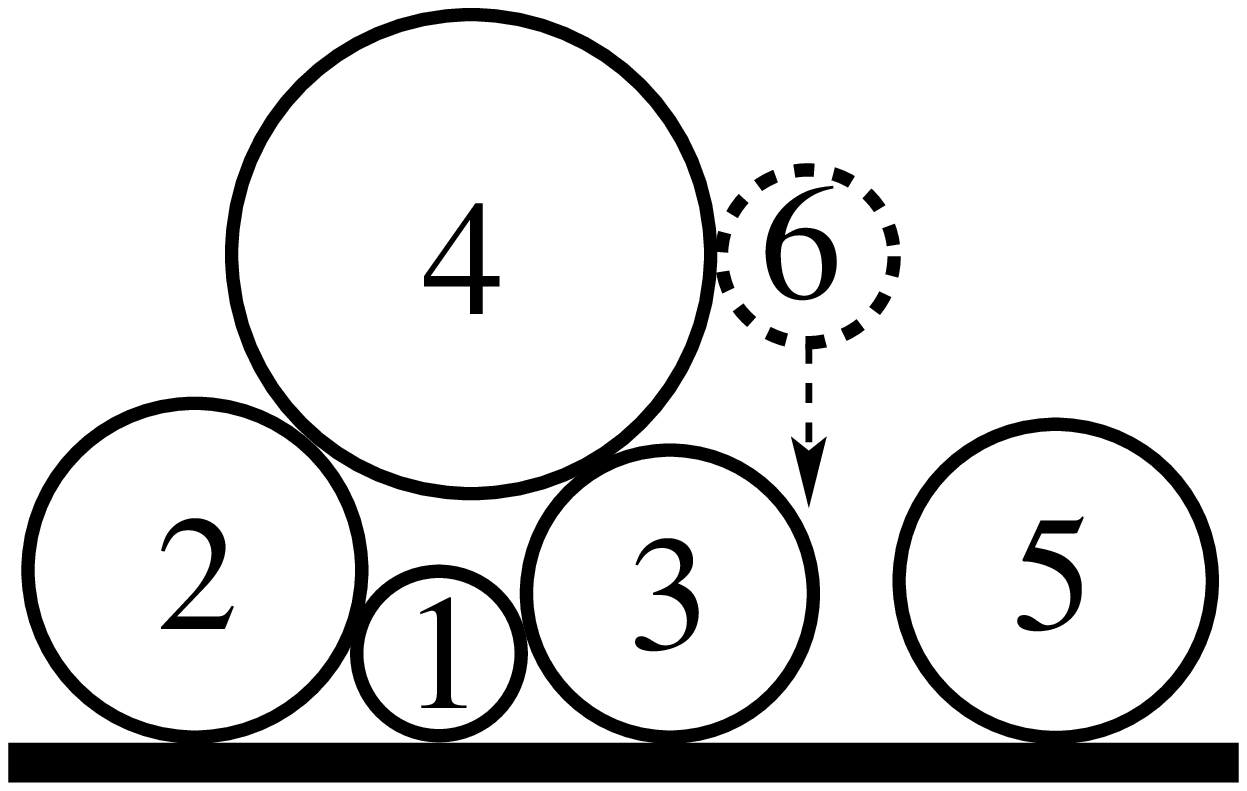}&\hspace*{\TSTPnegdist}
\includegraphics[width=\TSTPfigwidth]{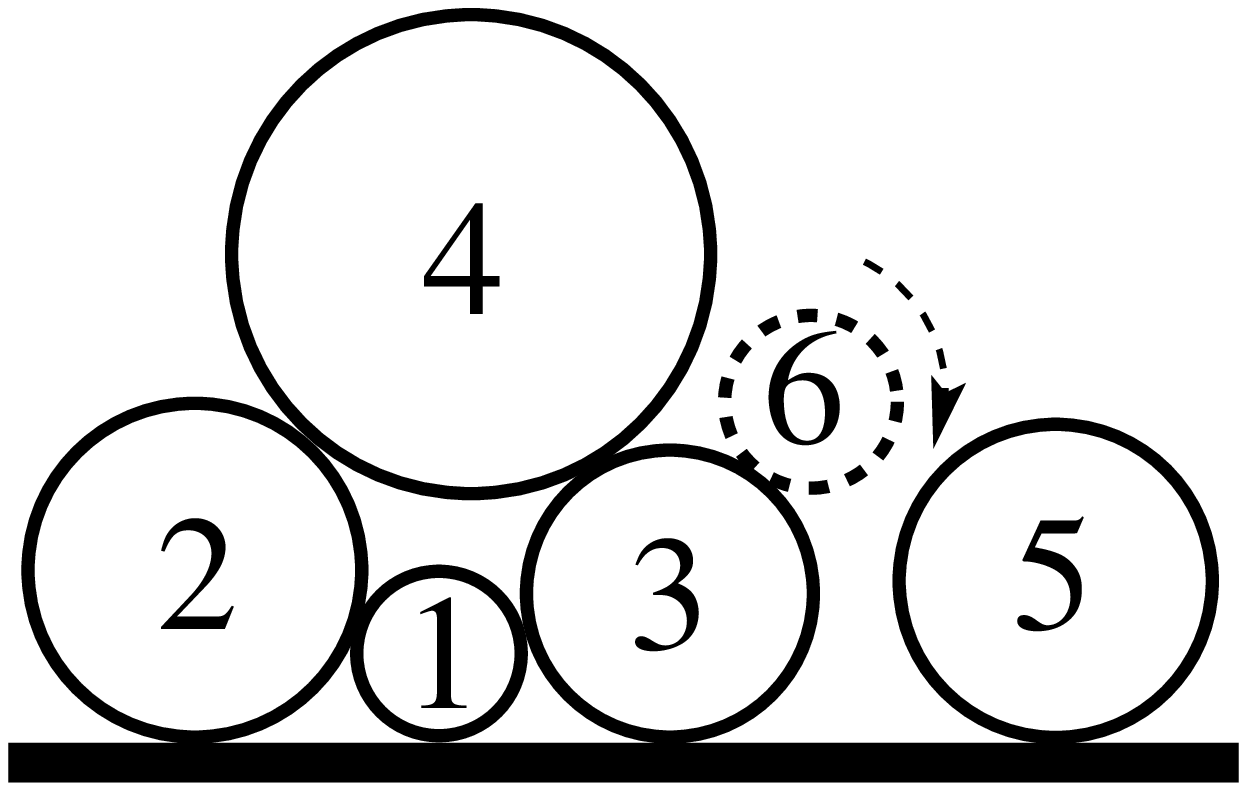}&\hspace*{\TSTPnegdist}
\includegraphics[width=\TSTPfigwidth]{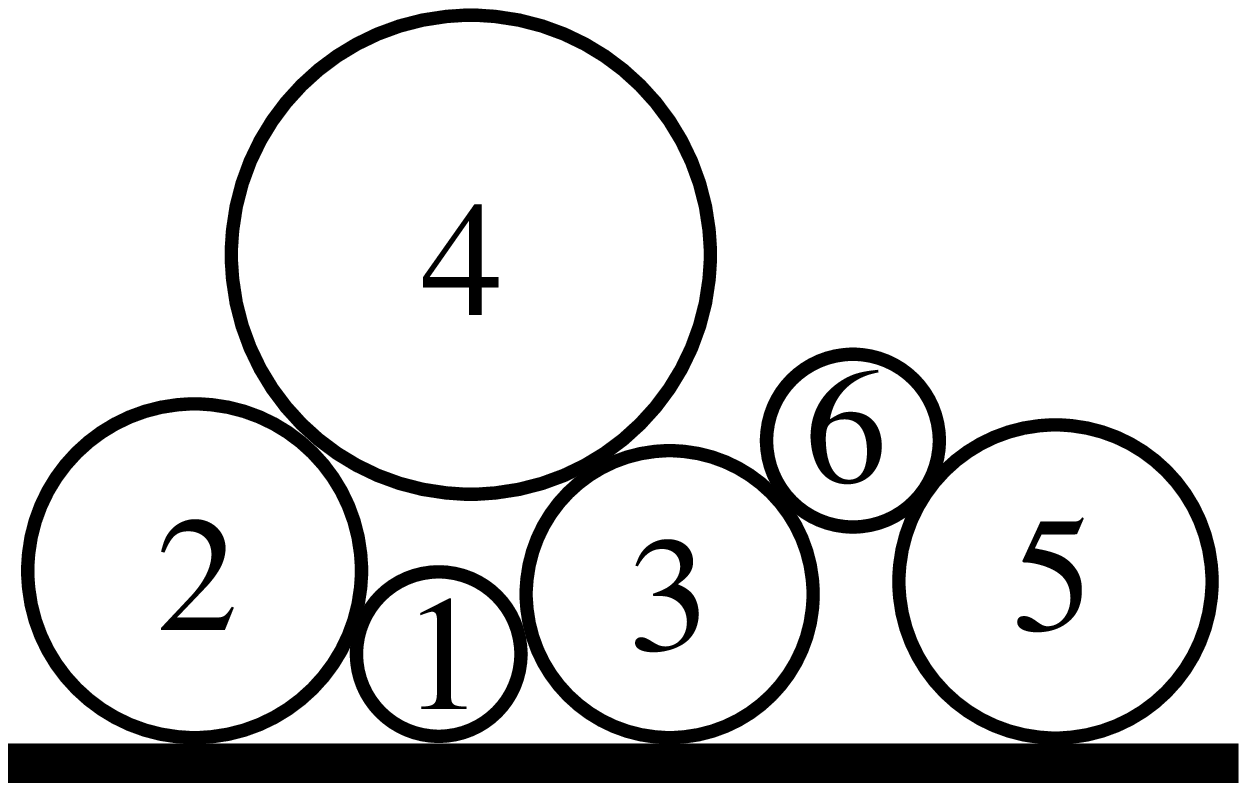}
\\
\multicolumn{5}{p{\columnwidth}}{\footnotesize Particle 6 first touches particle 4 and rolls on it until it loses contact when the centers of both particles are at the same height. Then the particle falls again until it touches particle 3. It continues rolling until it touches particle 5 too where it is fixed (position is stable since it is a local minimum).}
\end{tabular}
\caption{Sketch of the BTR algorithm}
\label{fig:sketch}
\end{figure}

In Fig. \ref{fig:sketch} the algorithm is sketched for a two-di\-men\-sio\-nal system. The
moving particle is drawn with a dotted line, fixed particles are
drawn with a solid line.

The algorithm by Visscher and Bolsterli \cite{VisscherBolsterli:1972} was later improved and generalized, e.g., \cite{jullien87b,jullien88,jullien92} and also applied to certain dynamical systems \cite{baumann94,baumann95}, see below.

The BTR algorithm is not exact, that is, it does not solve the coupled set of Newton's equations for the dynamics of the many-particle systems. Instead, there are three major simplifying assumptions
\begin{enumerate}
\item The motion of each particle $i$ is computed with the wall and the other already-deposited particles $j=0,\dots,i-1$ considered as fixed obstacles. The positions of the particles $j=0\dots i-1$ are not influenced by the motion of the new particle $i$. Thus, the trajectory of each particle is computed while taking gravity as the only driving force into account. The other particles $j=0\dots i-1$ and the wall establish the (complicated) boundary conditions to this motion. This way, the system of Newton's equation for the $N$-body system is decoupled and $N$ single-particle equations are solved instead. 
\item Collisions of particle $i$ with the wall or with other particles are assumed to be perfectly inelastic.
\item The time dependence of the particles' motion is disregarded. 
\end{enumerate}
These simplifications increase the efficiency of the simulation significantly. Jullien et al. \cite{jullien93b} were able to simulate packings of up to $10^8$ particles using this algorithm. The simplifications restrict the applicability of BTR to a rather small class of problems, see Sec. \ref{sec:bench}. If applicable, however, BTR is a very efficient simulation tool.

\section{Implementation of the BTR algorithm}
\label{sec:VisscherHeap}

The BTR algorithm allows for a very efficient numerical simulation. We briefly explain the implementation for the simulation of a two-dimensional heap which is the simplest application of BTR and restrict ourselves to the simulation of a two-dimensional system of polydisperse particles. The extension to three dimensions is straightforward. A more detailed explanation can be found in \cite{algo} and the full source code of our program is available at \cite{source}.

To compute the deposition of the $n$th particle, we do not need the positions of all $n-1$ previously deposited particles. Instead, it is sufficient to know the positions of particles that can come into contact with particle $n$, i.e., particles at the surface of the heap. This simplification is always justified for two-dimensional systems, whereas for three-dimensional systems it is only justified if the ratio between the largest and smallest radii of the particles in the system does not exceed the Apollonian ratio
$r_A\equiv R_{\rm max}/R_{\rm min} =
\sqrt{3}\left/\left(2-\sqrt{3}\right)\right.\approx 6.46$. Otherwise the smaller particles can fall into the gaps between the larger particles and thus, in principle, interact with all particles of the heap. In this case, we cannot exclude any pair interactions \`a priori.
A natural choice of radii within $(R_{\rm min}, R_{\rm max})$ is to assume
that the total mass of particles from the interval $(R,R+{\rm d} R)$ is constant regardless of $R$. Such a distribution can be obtained from equi-distributed random numbers $z\in[0,1)$ using the transformation \cite{algo,PressVetterlingTeukolskyFlannery:1988}
\begin{equation}
  R_i=\frac{R_{\rm min} R_{\rm max}}{R_{\rm max}-z\left( R_{\rm max}-R_{\rm min}\right)}\;.
\end{equation}

We store the indices of the particles at the surface in the container \verb|surface| of type \verb|map<double,int>| where the key is the their $x$-coordinate. 

\begin{figure}[h!]
  \centering
  \includegraphics[width=4.5cm]{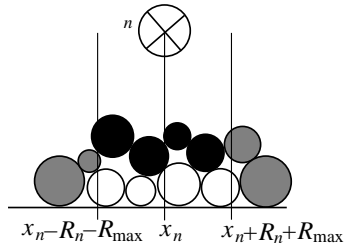}
  \caption{The centers of the contact candidates of the falling particle $n$ (crossed circle) belong to the interval $(x_n-R_n-R_{\rm max}, x_n+R_n+R_{\rm max})$ (black filled circles). The other surface particles are drawn gray, particles which are not part of the surface are drawn hollow}
  \label{fig:fall}
\end{figure}

Consider the deposition of particle $n$. If $n$ is not in contact with any other particle it moves downward until it touches the ground or another particle. Apart from the ground plane all particles on the surface of the heap whose centers belong to the interval $(x_n-R_n-R_{\rm max}, x_n+R_n+R_{\rm max})$ are potential contact partners, see Fig.~\ref{fig:fall}. The container \verb|map| allows to very efficiently iterate through these contact candidates. 

\begin{figure}[h!]
  \newcommand{\TSTPfigwidth}{2.1cm}
  \newcommand{\TSTPnegdist}{-0.51cm}
  \centering
  \includegraphics[width=\TSTPfigwidth]{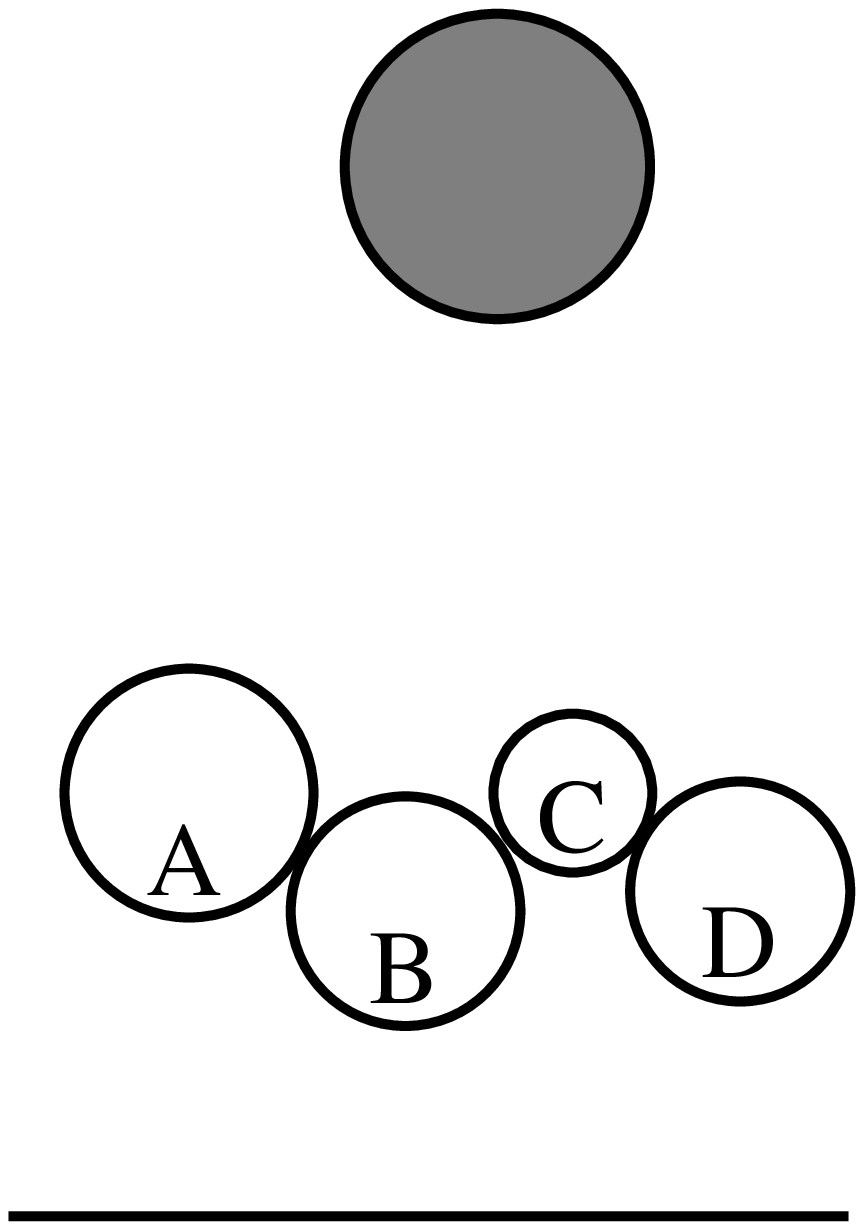}\hfill
  \includegraphics[width=\TSTPfigwidth]{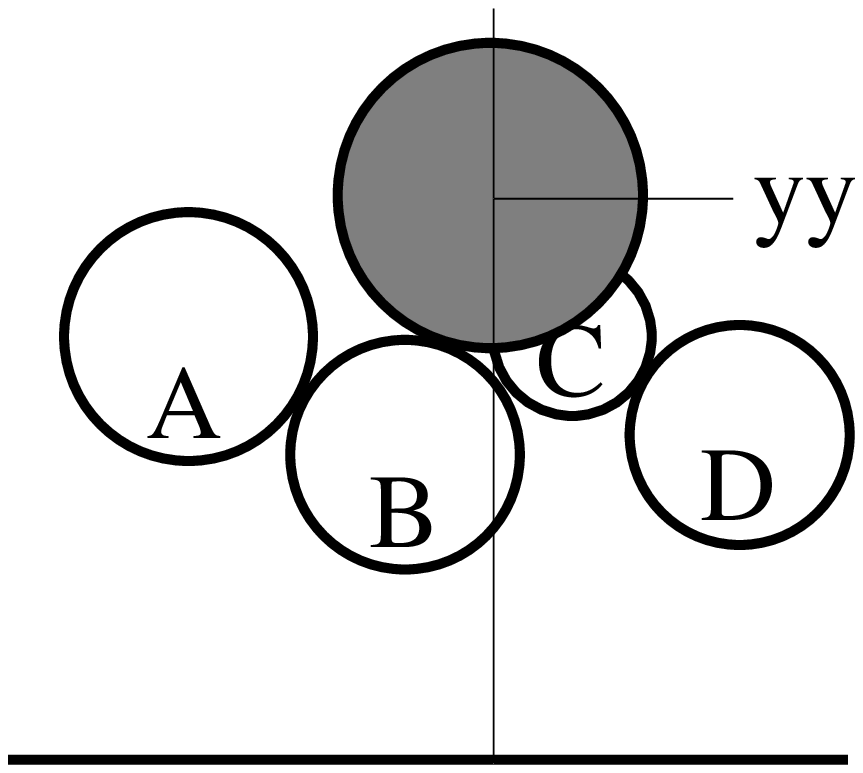}\hfill
  \includegraphics[width=\TSTPfigwidth]{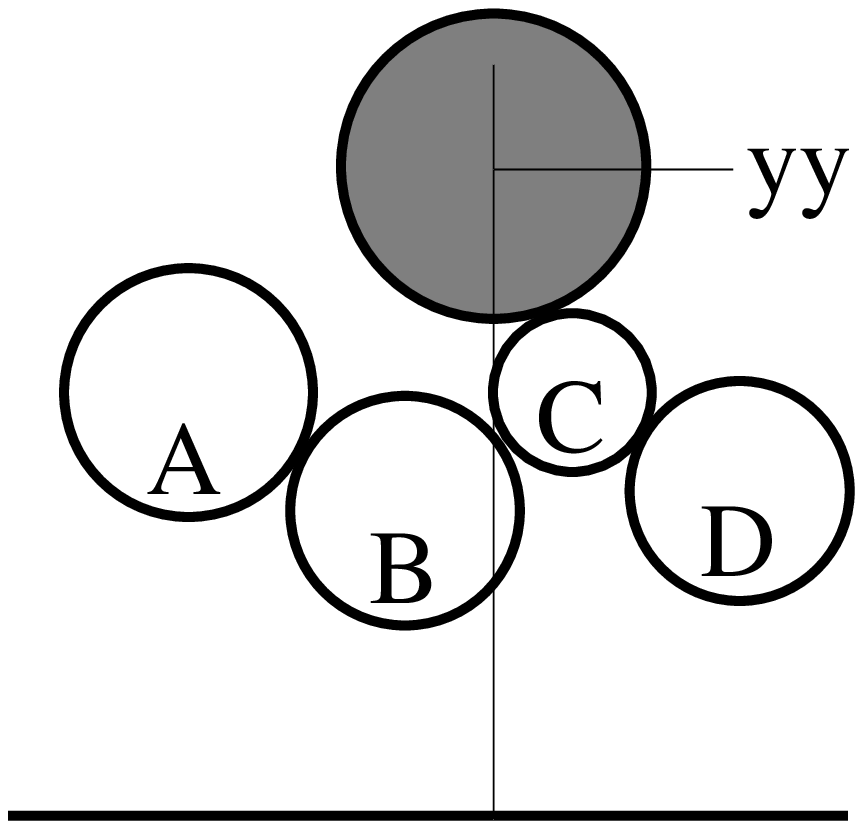}\hfill
  \includegraphics[width=\TSTPfigwidth]{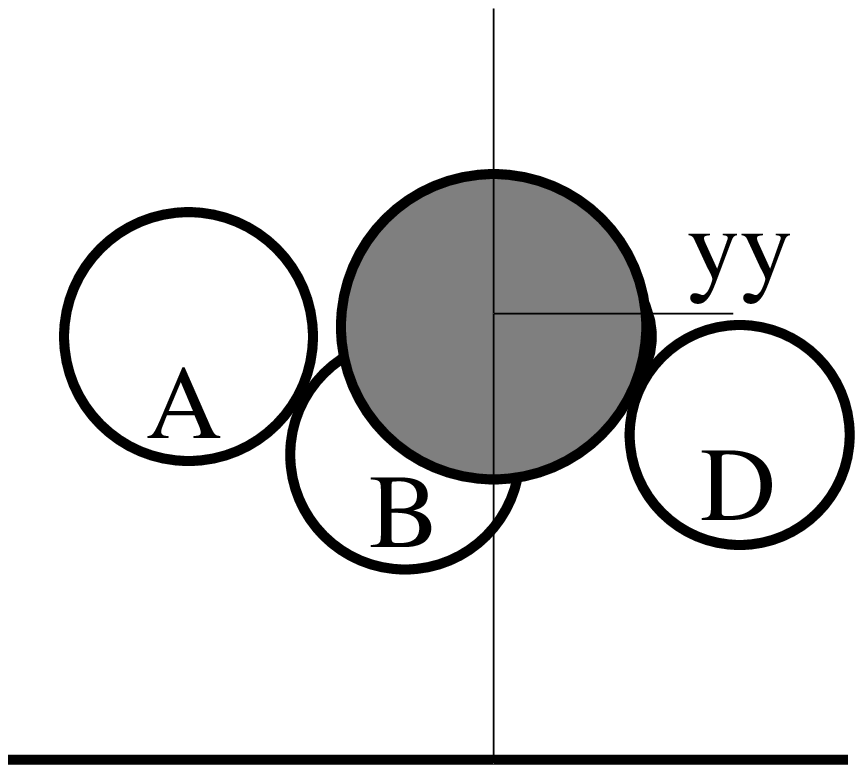}
  \caption{For each of the candidates $A$-$D$ (drawn in black in Fig. \ref{fig:fall}) we determine the vertical coordinate $yy$ of a possible contact. The highest contact is with candidate $C$, therefore, $C$ is the contact partner we are looking for}
  \label{fig:fallx}
\end{figure}

For all contact candidates $i$ that are located below the particle to be deposited, we check if and at which height $yy$ both particles touch, disregarding for the moment possible interference from other particles, that is, we check whether 
\begin{equation}
  {yy} = y_i+\sqrt{(R_n+R_i)^2-(x_i-x_n)^2}\;,
\end{equation}
has a real solution and determine the maximum of this expression for all $i$ (see Fig. \ref{fig:fallx}). To determine the desired contact point also a possible contact with the ground must be taken into account.
Particle $n$ is now moved downward to contact its partner particle or the ground. In the latter case, its new position is stable by definition and its deposition is accomplished. If, however, the particle meets another particle first, its new position cannot be stable. Instead it continues its motion by rolling down the surface of the contact partner until it comes into contact with the wall or with another particle, or it continues to fall down, as sketched in Fig. \ref{fig:sketch}.

\begin{figure}[h!]
  \newcommand{\TSTPfigwidth}{2.8cm}
  \centerline{
    \hfill\includegraphics[height=\TSTPfigwidth]{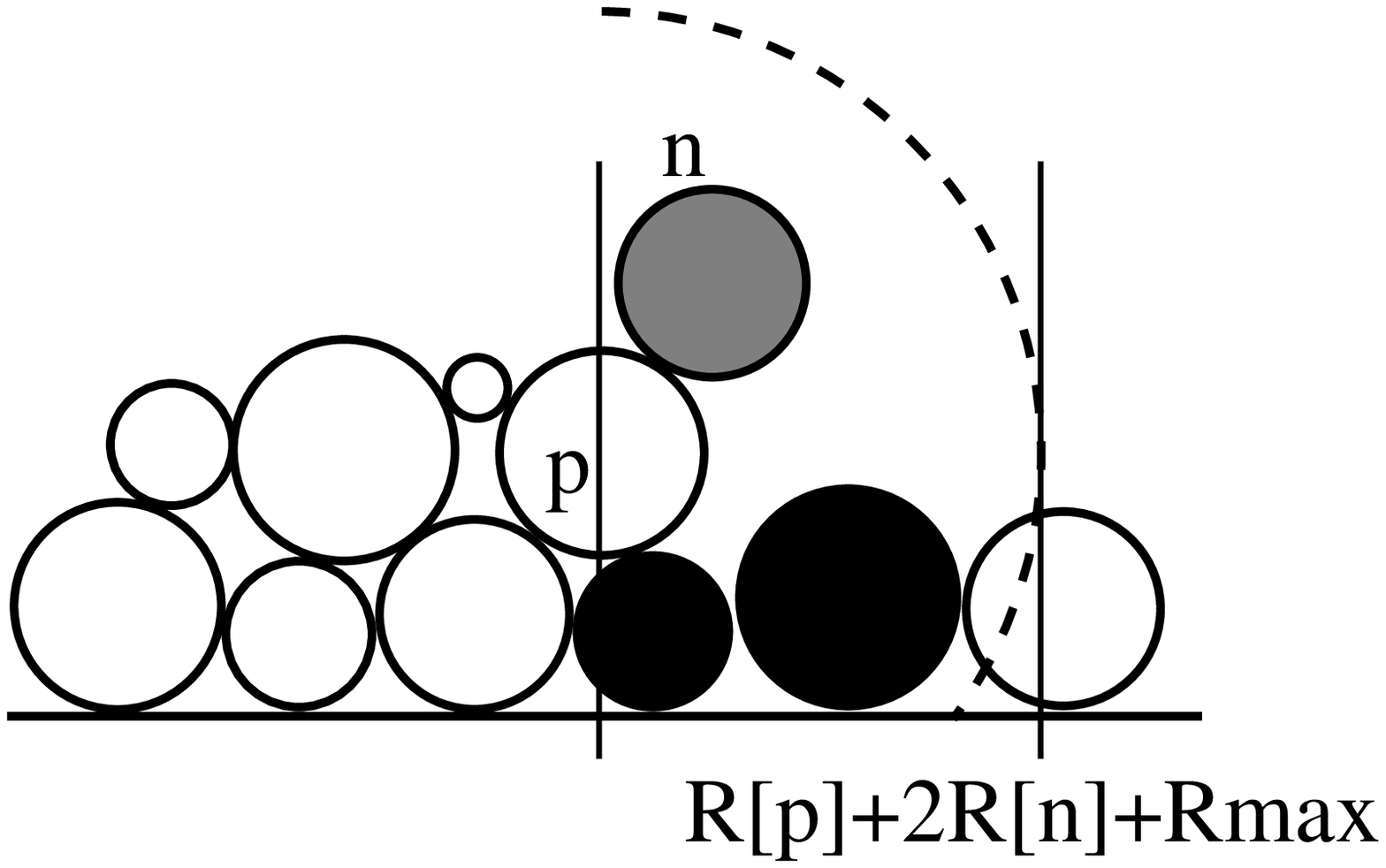}
    \hfill\includegraphics[height=\TSTPfigwidth]{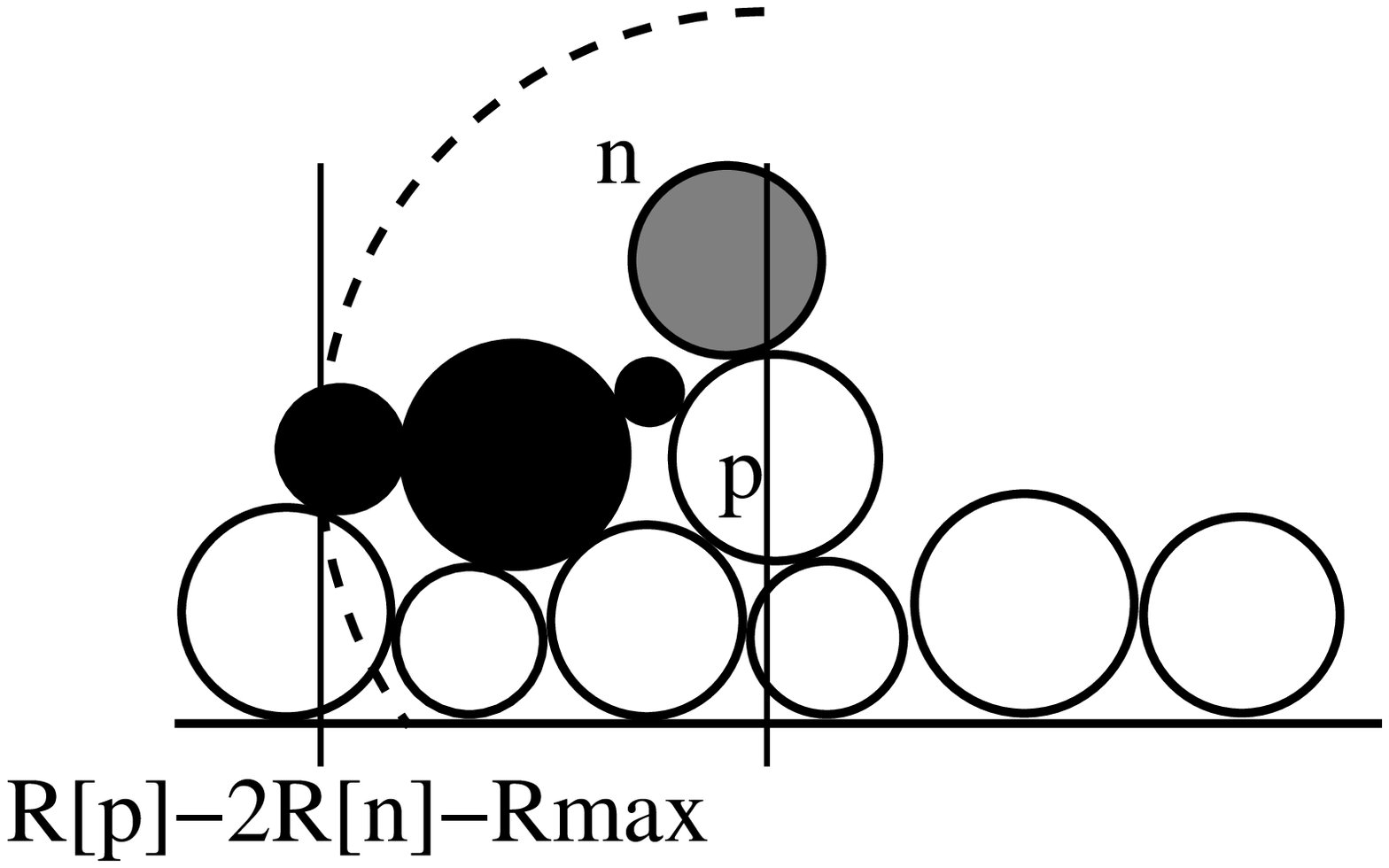}
    \hfill}
  \caption{The centers of the candidates for the second partner
    particle are located within the circle of radius
    $r=R_p+2R_n+R_{\rm max}$ around $x_p$. Depending on the relative
    position of the particles $n$ and $p$ the circle can be reduced to a half circle. The particles represented by filled circles are the candidates}
  \label{fig:findX}
\end{figure}

Similar as described above, again all possible candidates for the additional contact partner are determined, i.e., all particles that may be contacted by particle $n$ while maintaining contact to its present partner $p$. The new particle rolls to the left if $x_n<x_p$ or to the right otherwise. Candidates for the contact are particles whose $x$-coordinate is in the interval $x_p\dots x_p\pm (R_p+2R_n+R_{\rm max})$, see Fig. \ref{fig:findX}. Again the index range of the candidate partners can be obtained easily from the map \verb|surface|. 

We investigate the possible contact of particle $n$ rolling down the surface of particle $p$
with each candidate $i$ separately, disregarding all other particles for the
moment, that is, we seek real solutions $\vec{r}_n^{\,i}\equiv (x_n^i,y_n^i)$ of the system
\begin{equation}
  \begin{split}
    \left|\vec{r}_n^{\,i}-\vec{r}_{p}\right|&= R_n+R_{p}\\
    \left|\vec{r}_n^{\,i}-\vec{r}_{i}\right|&= R_n+R_{i}\,,
  \end{split}
  \label{eq:pcontact}
\end{equation}
for contacts between particles $n$, $p$ and $n$, $i$ and of the system
\begin{equation}
  \begin{split}
    \left|\vec{r}_n^{\,w}-\vec{r}_{p}\right|&= R_n+R_{p}\\
    y_n^w = R_{n}\,.
  \end{split}
  \label{eq:pcontact1}
\end{equation}
for a possible contact between particles $n$ and $p$ and of $n$ with the floor. If
these systems have real solutions the new position of particle $n$ is
determined by the solution $\vec{r}_n^{\,i}\equiv (x_n^i,y_n^i)$ or
$\vec{r}_n^{\,w}\equiv (x_n^w,y_n^i)$ with the largest
maximum vertical component, provided, $y_n^i$ or respectively $y_n^w$ is
larger than $y_p$. In this case we found a new position of particle
$n$. We now check for stability (i.e. if the particle is supported both from the right of his center and from the left). If it is stable the deposition of the particle is finished, otherwise particle $i$ assumes the role of particle $p$ and particle $n$ continues to roll on its surface.
If no candidate is found before particle $n$ reaches $y_n=y_p$, it loses contact with $p$ and again falls
vertically downward, see last line in Fig. \ref{fig:sketch}. 
This procedure is repeated until a stable position of particle $n$ is found.

\begin{figure}[h!]
  \newcommand{\TSTPfigwidth}{2.9cm}
  \newcommand{\TSTPnegdist}{-0.51cm}
  \centerline{
    \hfill\includegraphics[width=\TSTPfigwidth]{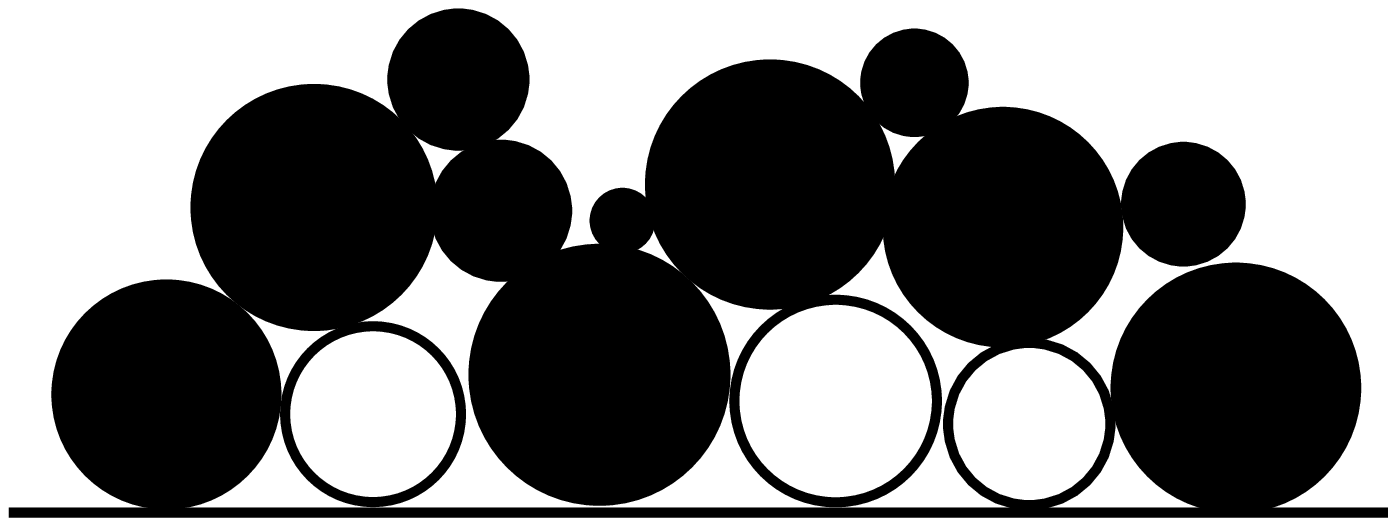}
    \hfill\includegraphics[width=\TSTPfigwidth]{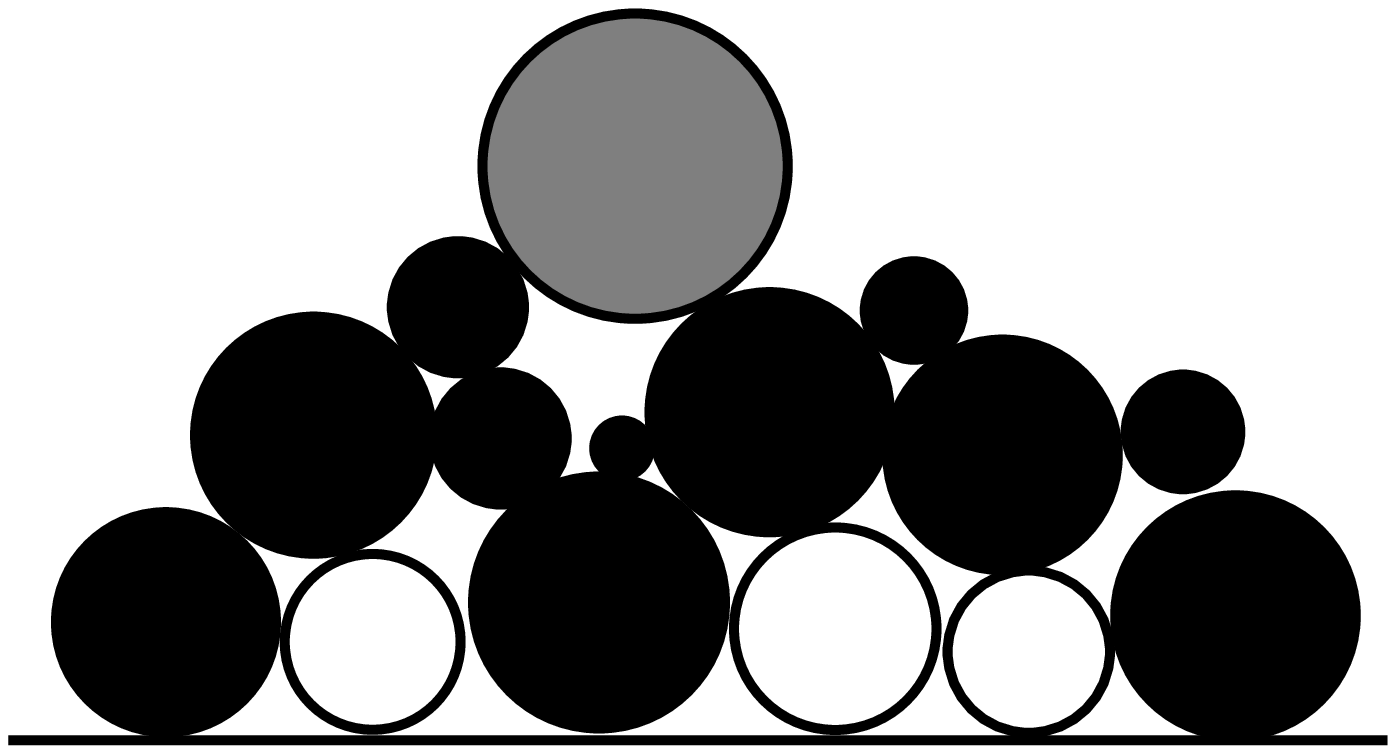}
    \hfill\includegraphics[width=\TSTPfigwidth]{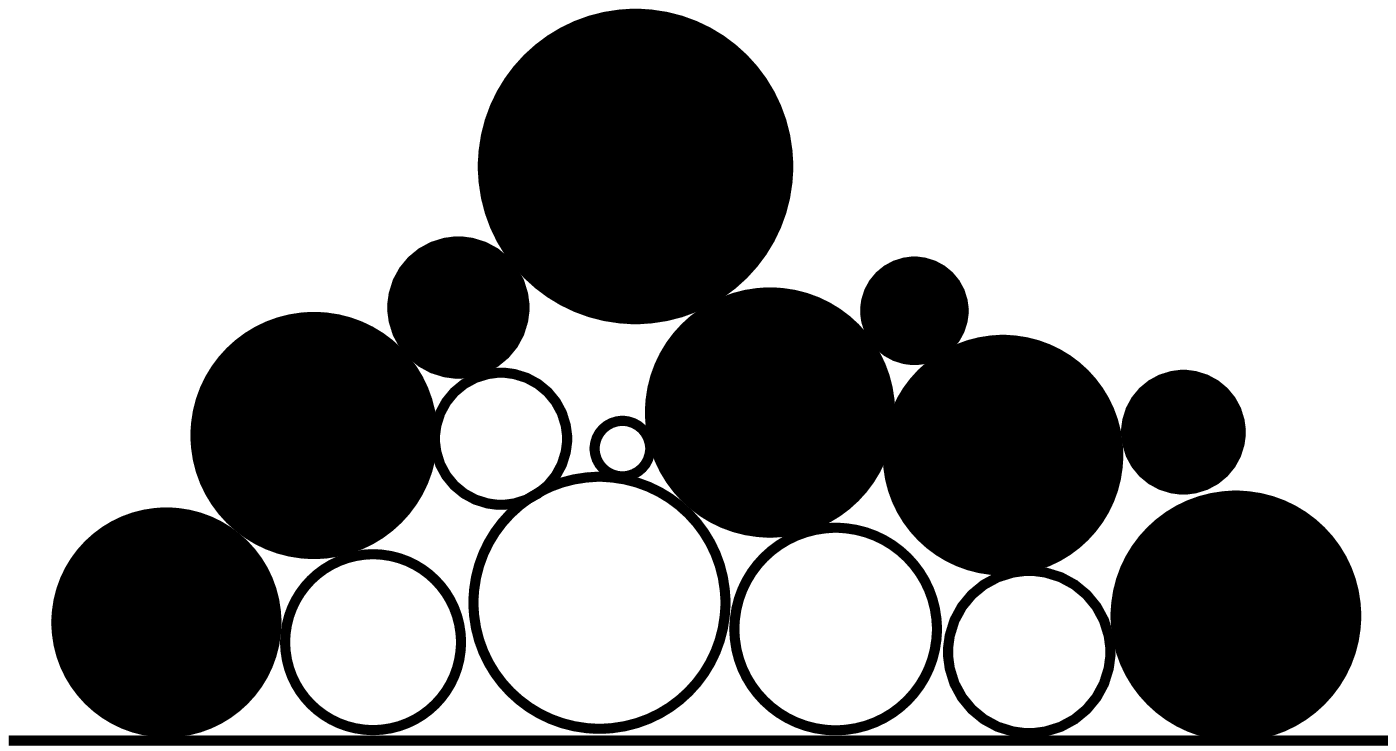}
    \hfill}
 \caption{\emph{Left}: surface particles before the deposition of the
 new particle (drawn gray).  \emph{Center}: a new particle is deposited. \emph{Right}: corrected list of surface particles}
  \label{fig:updatesurface}
\end{figure}

When the stable position of $n$ is found, the map \verb|surface| is
updated. First, $n$ is recorded as a member of the map since it became part of the surface.
Second, after depositing $n$, other particles may be screened, such that no
further particles may come into contact with them. Consequently, these
particles are not members of the surface anymore and are removed from
\verb|surface|, see Fig. \ref{fig:updatesurface}.

\begin{figure}[h!]
  \centering
  \includegraphics[height=3.8cm]{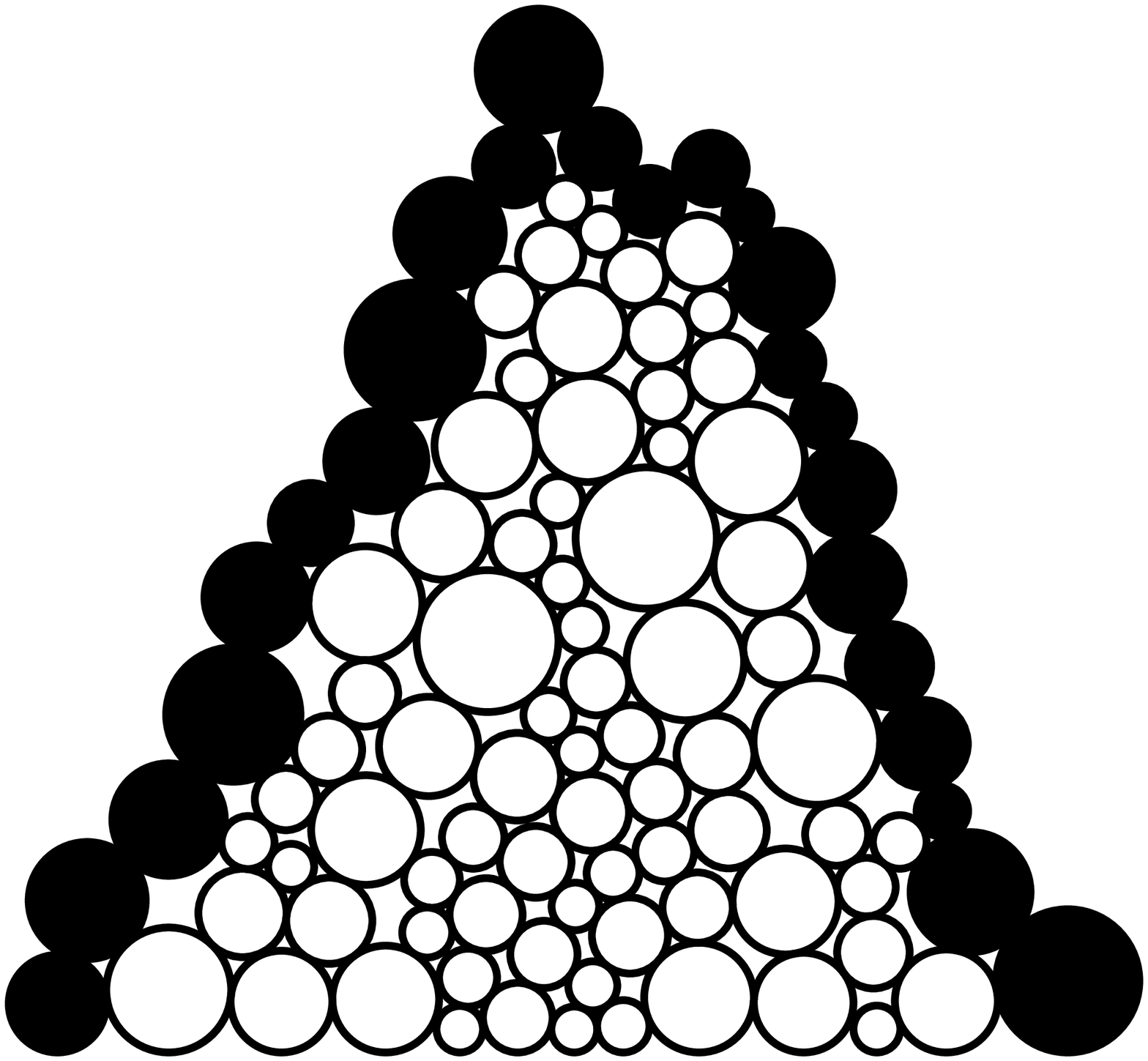}~
  \includegraphics[height=3.4cm]{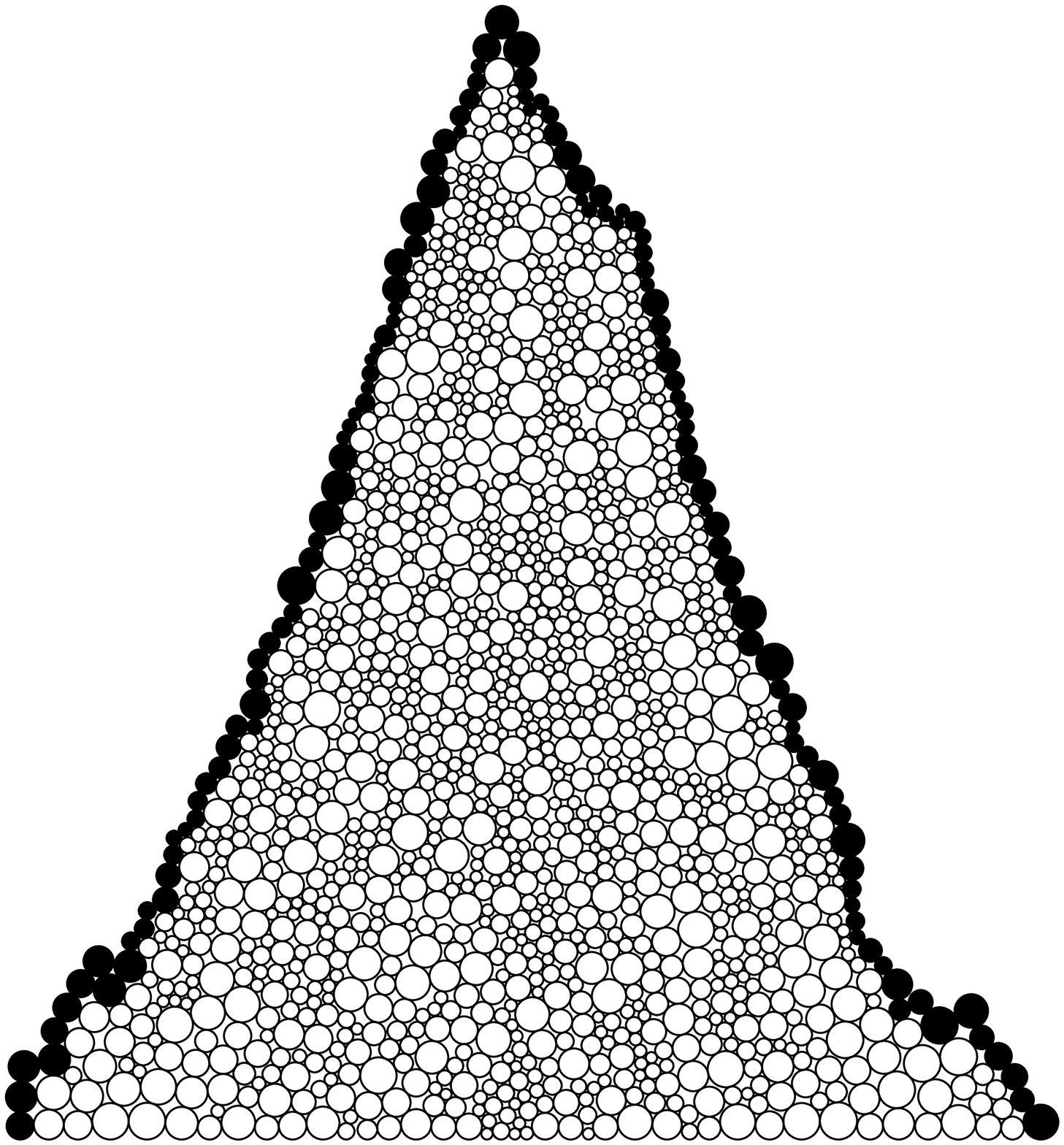}
  \caption{Snapshots of the growing heap. The number of particles is $N=100$ and $N=1400$}
  \label{fig:Visscher.heap}
\end{figure}

The description of the algorithm and the sketch of the implementation is now
complete. Figure \ref{fig:Visscher.heap} shows snapshots of a growing
heap. The particles which are part of the surface are drawn filled. 

Since particles in the process of deposition can contact only the particles at the surface, unlike as in
MD the computational complexity of depositing a particle $n$ does not grow 
with the number of already deposited particles, ${\cal O}(n)$ but it increases
with ${\cal O}(n^{1/2})$ in 2d or with ${\cal O}(n^{2/3})$ in 3d which is the
reason for the computational power of BTR. We discuss the efficiency of the
algorithm below in Sec. \ref{sec:bench}.

\section{Example: Stratification in a Sand Heap}
\label{sec:stratification}

\begin{figure}[h!]
  \centering
  \includegraphics[height=3.2cm,clip]{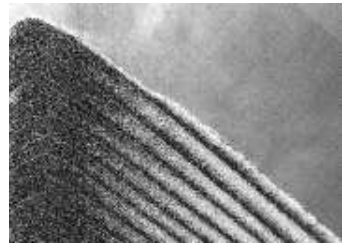}
  \caption{Formation of stripes in a heap consisting of particles of different properties (experimental result). The figure was taken from 
    \cite{gray00:_shock}}
  \label{fig:Visscher.Exp}
\end{figure}

When a heap of particles is created by sequentially depositing particles of
different size, size segregation (stratification) is observed, which is caused
by different angles of repose for large and small particles
\cite{LitwiniszynCiTong:1963,Williams:1976,Brown:1939,Williams:1963}. The
particles form stripes as shown in Fig. \ref{fig:Visscher.Exp}. 
In three dimensional dunes and ripples, other more complex stratification patterns are observed \cite{Allen:1962}. This effect has been studied and modeled extensively, e.g., in \cite{KoeppeEnzKakalios:1997,JulienLanRaslan:1997,MakseHavlinKingStanley:1997,Makse:1997,MakseHavlinKingStanley:1997a,MakseHavlinKingStanley:1997,MakseEtAl:1996,MakseCizeauStanley:1997,MakseCizeauStanley:1998,MakseHerrmann:1997,GrayHutter:1997,GrayHutter:1998,GrasselliHerrmann:1997a,BoutreuxDeGennes:1997,BoutreuxDeGennes:1996,Meakin:1990}. Similar structures have been observed in sand overflown by wind or water \cite{Sorby:1859}. 

\begin{figure}[h!]
\centerline{\includegraphics[width=\columnwidth,height=7cm]{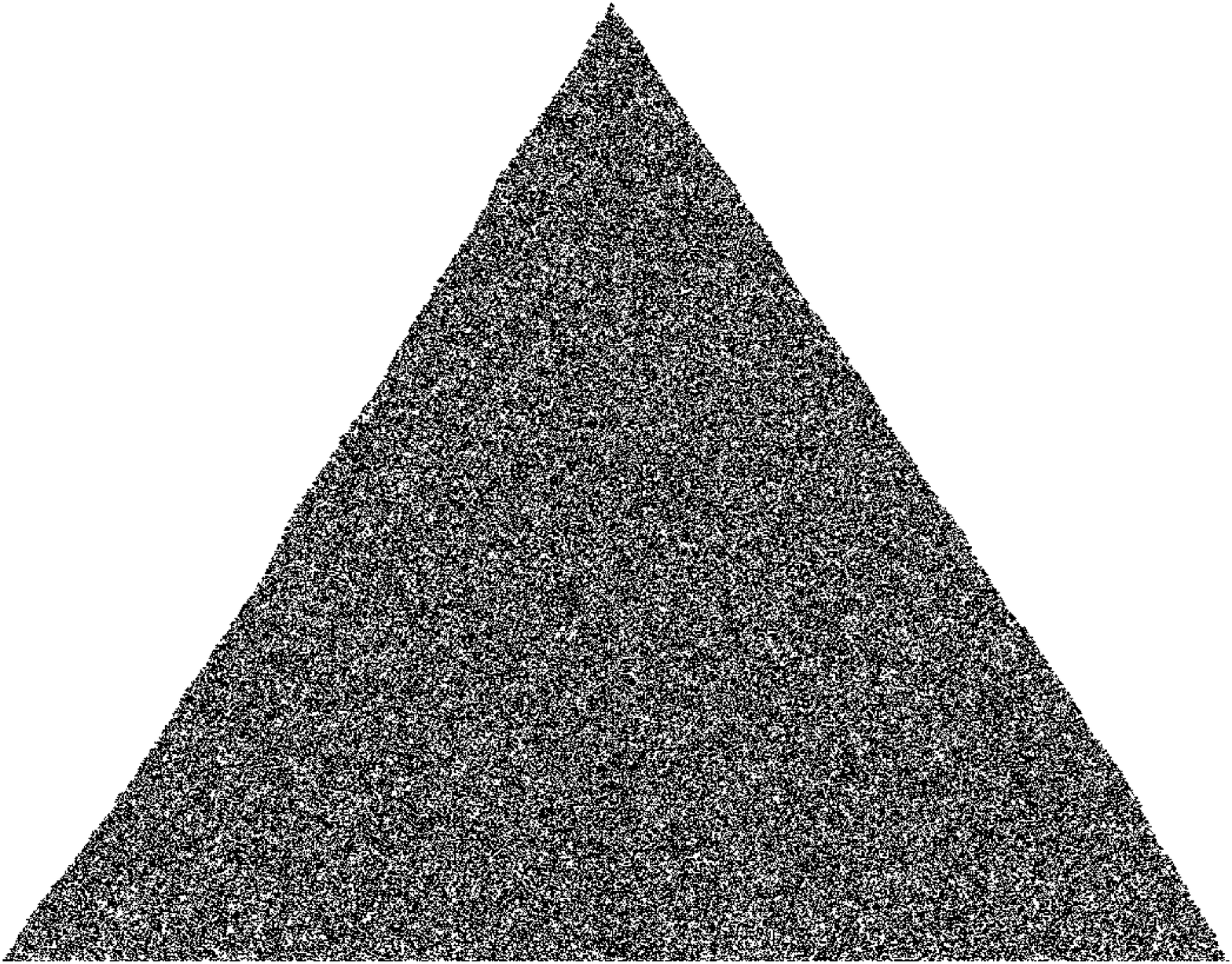}}
\vspace*{0.2cm}
\centerline{
  \includegraphics[width=0.48\columnwidth,height=0.48\columnwidth]{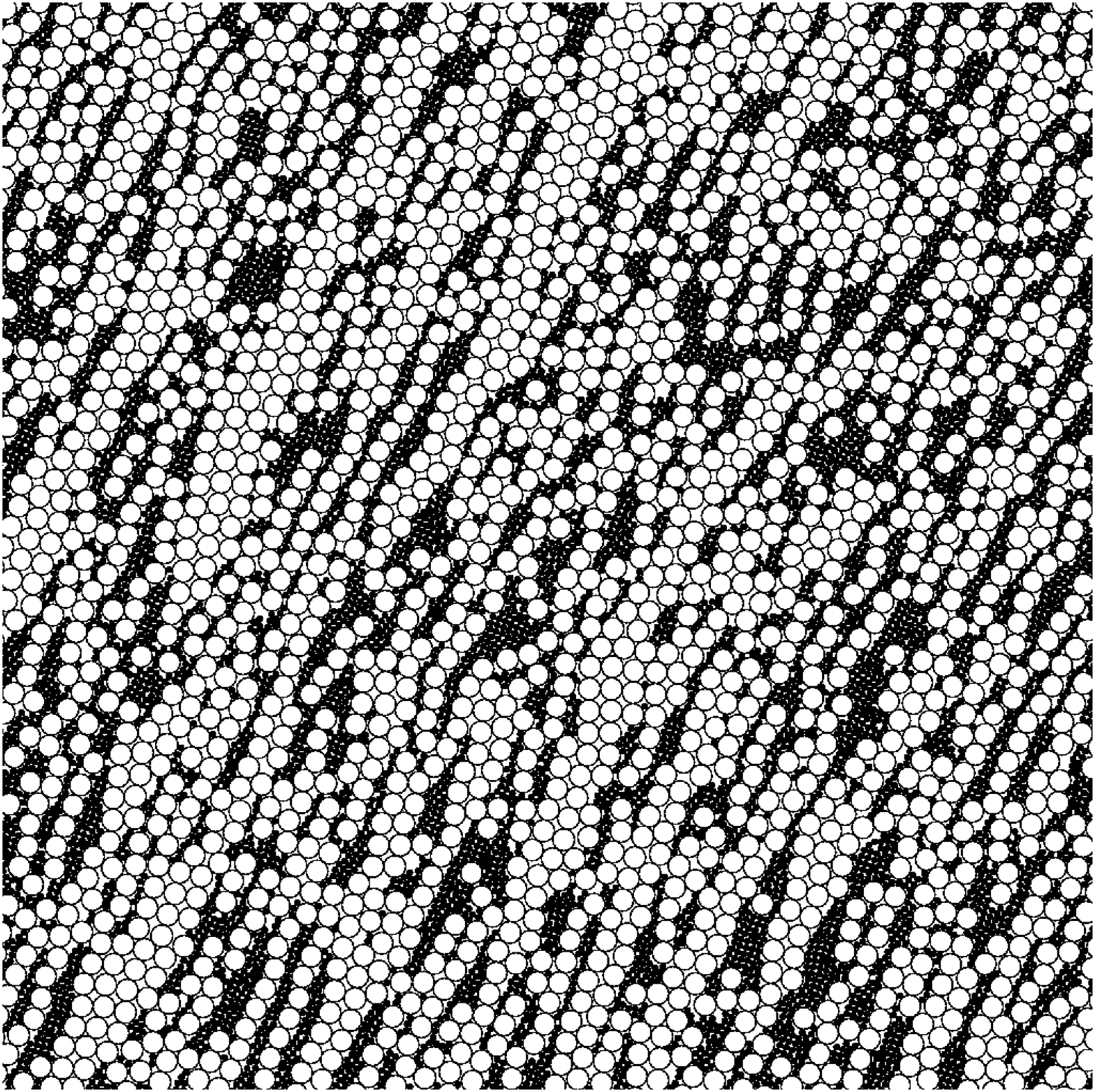}~
  \includegraphics[width=0.48\columnwidth,height=0.48\columnwidth]{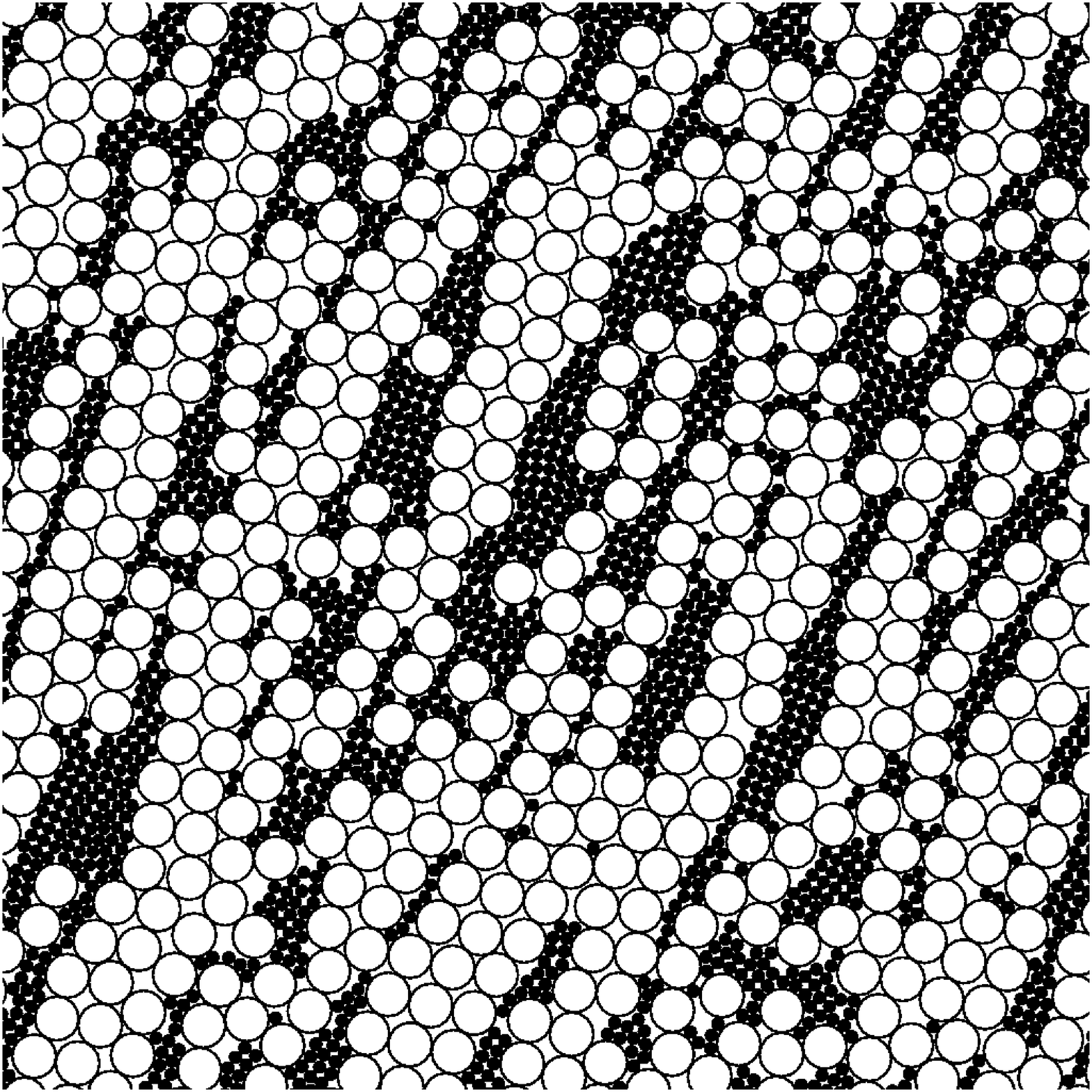}  
}
  \caption{A heap consisting of $N=10^6$ particles of two different radii. Size segregation (stratification) can be seen in the close-ups}
 \label{fig:Visscher.Stratifikation}
\end{figure}

Figure \ref{fig:Visscher.Stratifikation} shows a heap of $N\!=\!10^6$
particles of two different radii. The effect of stratification is
visible in the close-ups. The small particles are drawn filled.  


\section{Sedimentation of Nano-Powders}

The BTR-method can be generalized to clusters of contacting particles. Here clusters are
considered as perfecly rigid, that is, the particles belonging to a cluster do
not change their relative positions. This idealized behavior may be adequate
for cohesive nano-powders where the attractive surface forces exceed by far inertial
forces, due to their enourmous surface area per unit mass. The attracting force between particles in contact provides a mechanism which fixes particles to a position where they where first deposited. Also due to
the surface forces, nano-powders are frequently very pourous. In this section
we describe briefly the application of BTR to coarsening of nano-powders in
the process of repeated siphoning the material from one container into
another. For a more detailed description see \cite{SchwagerPoeschel:inprep}. 

For clusters, the first step of the BTR algorithm is handled as for simple 
spheres: We determine the position of the falling cluster when one of its
particles touches a particle of the sediment or the wall.

Rolling of the cluster on the already deposited heap particles is more
complicated. Apart from situations discussed in Sec. \ref{sec:VisscherHeap} 
one is confronted with a large number of special cases whose discussion 
is outside the scope of this work \cite{SchwagerPoeschel:inprep}. 

\begin{figure}[htbp]
  \centerline{\includegraphics[width=0.5\columnwidth]{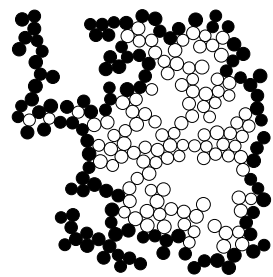}}
  \caption{Typical cluster after a few disposition cycles. Dark circles represent surface particles. The surface is not contiguous, the gaps are, however, too small for outside particles to touch inner particles.}
  \label{fig:clusters}
\end{figure}

\begin{figure}[htbp]
  \centerline{
    \includegraphics[width=0.48\columnwidth]{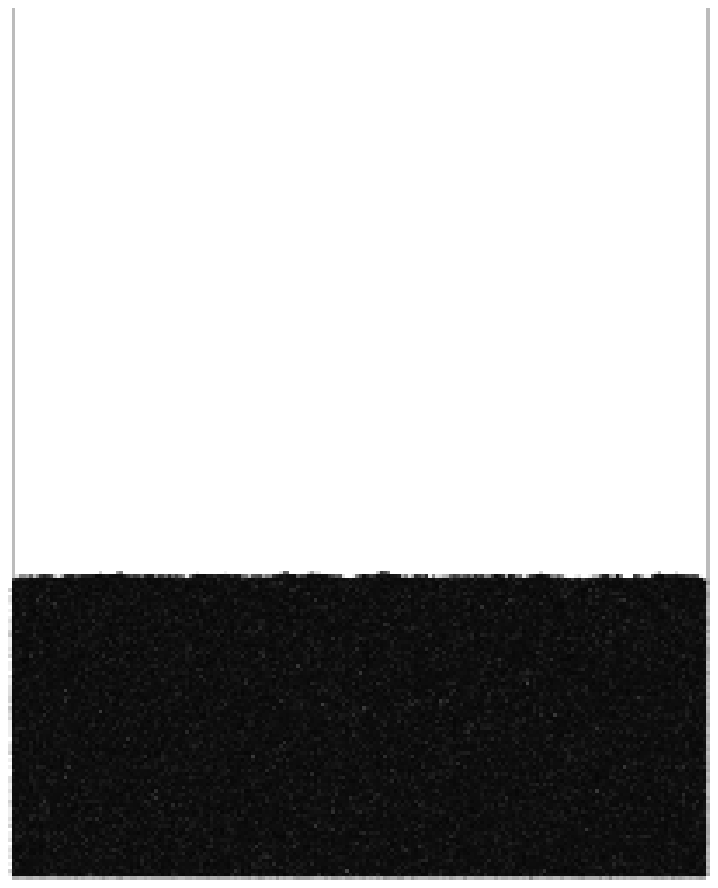}
    \includegraphics[width=0.48\columnwidth]{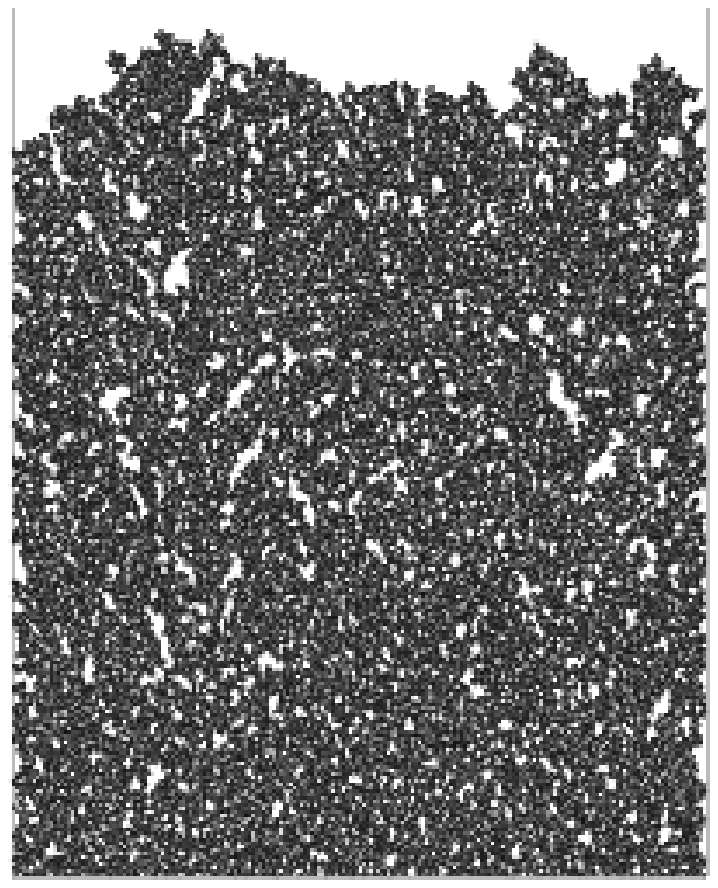}
  }
  \caption{A heap of 6 million particles. Left: Particles deposited in clusters of 3. Right: After 18 redeposition cycles.}
  \label{fig:clusterheap}
\end{figure}

\begin{figure}[htbp]
  \centerline{\includegraphics[width=0.8\columnwidth,clip]{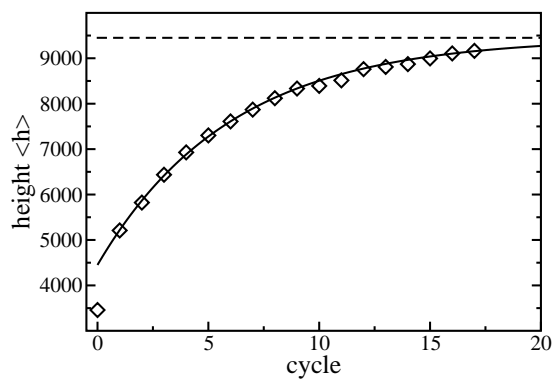}}
  \caption{Average height of the heap vs number of disposition cycles. The solid line is the fit to an exponential function, the diamonds are the measured values.}
  \label{fig:avgheight}
\end{figure}

For efficient simulation of both processes, falling and rolling, we need the
{\em accessible surface} of the cluster, that is, those particles which due to
geometry can come into contact with other particles or the floor. There are
several efficient algorithms to determine this surface \cite{ORourke}. 
Fig. \ref{fig:clusters} shows an example cluster and its surface (filled circles). The holes in the surface are small enough to keep outside particles away from the inner particles. 
In our simulation of a coarsening nano-powder of $N=6\cdot 10^6$ cohesive
particles, $R\in (0.9,1.1)$, 
initially the particles (or small clusters of a few particles) 
are placed into a rectangular container of length $L=8000$ limited by vertical walls.
After the initial sedimentation due to standard BTR they are densely packed in the container (see Fig \ref{fig:clusterheap}, left).
To obtain a flat surface of the heap one has to insert the particles at random
positions distribuited uniformly over the length of the box. The material is
now cut into square blocks of about $50\times 50$ average particle diameters. The blocks are decomposed into clusters of mutually contacting particles. Now these clusters are sedimented into the container. Figure \ref{fig:avgheight} shows the average height $\left<h\right>$ of the material surface over the number of sedimentation cycles.

This quantity was computed by cutting the box into narrow vertical strips and determining the height $h$ of the highest particle in each strip. The height of the strip is now $h+R$ where $R$ is the radius of the highest particle. Averaging over all strips yields $\left<h\right>$. The data points fit an exponential law of the form $\left<h\right>=9450-5000\exp(-n/6)$ with $n$ being the cycle number. The inital height does not fit into this scheme. This is due to the fact that the initial heap is build up from very small structures (clusters of size 3 in the example) while each later sedimentation cycle is done with structures of a typical size 50. Fig. \ref{fig:clusterheap} shows the heap after 18 sedimentation cycles. 

\section{Dynamic Simulations}
\label{sec:BTTRdynamics}

BTR cannot be applied directly to the simulation of dynamic processes since
due to the main principle of this algorithm, deposited particles cannot leave
their positions anymore. A very restricted class of dynamical systems can be
simulated, however, if we partition the dynamics into alternating steps of
collective motion and sequential deposition \cite{jullien92}. 

\begin{figure}[h!]
  \centerline{
  \includegraphics[width=0.25\columnwidth]{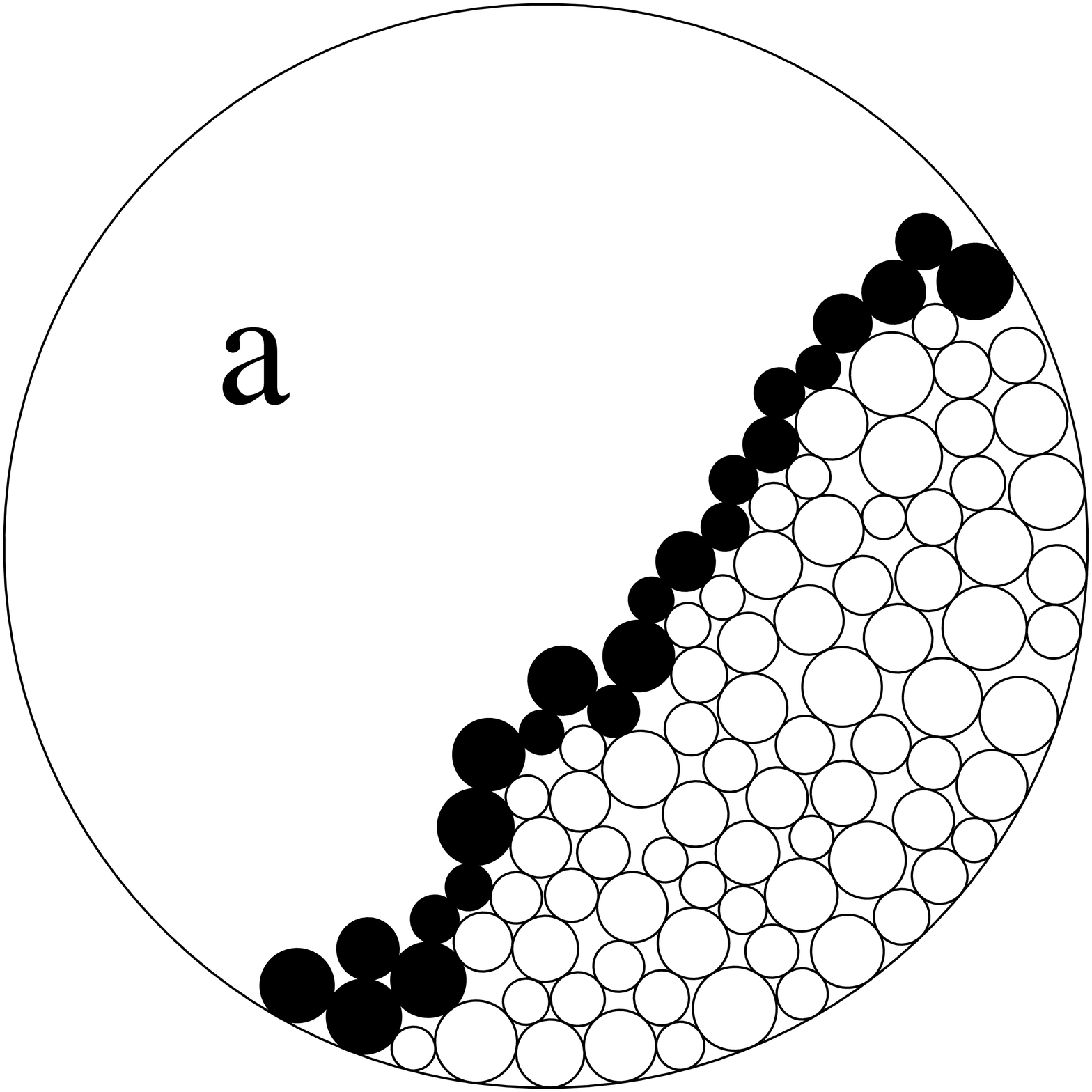}~
  \includegraphics[width=0.25\columnwidth]{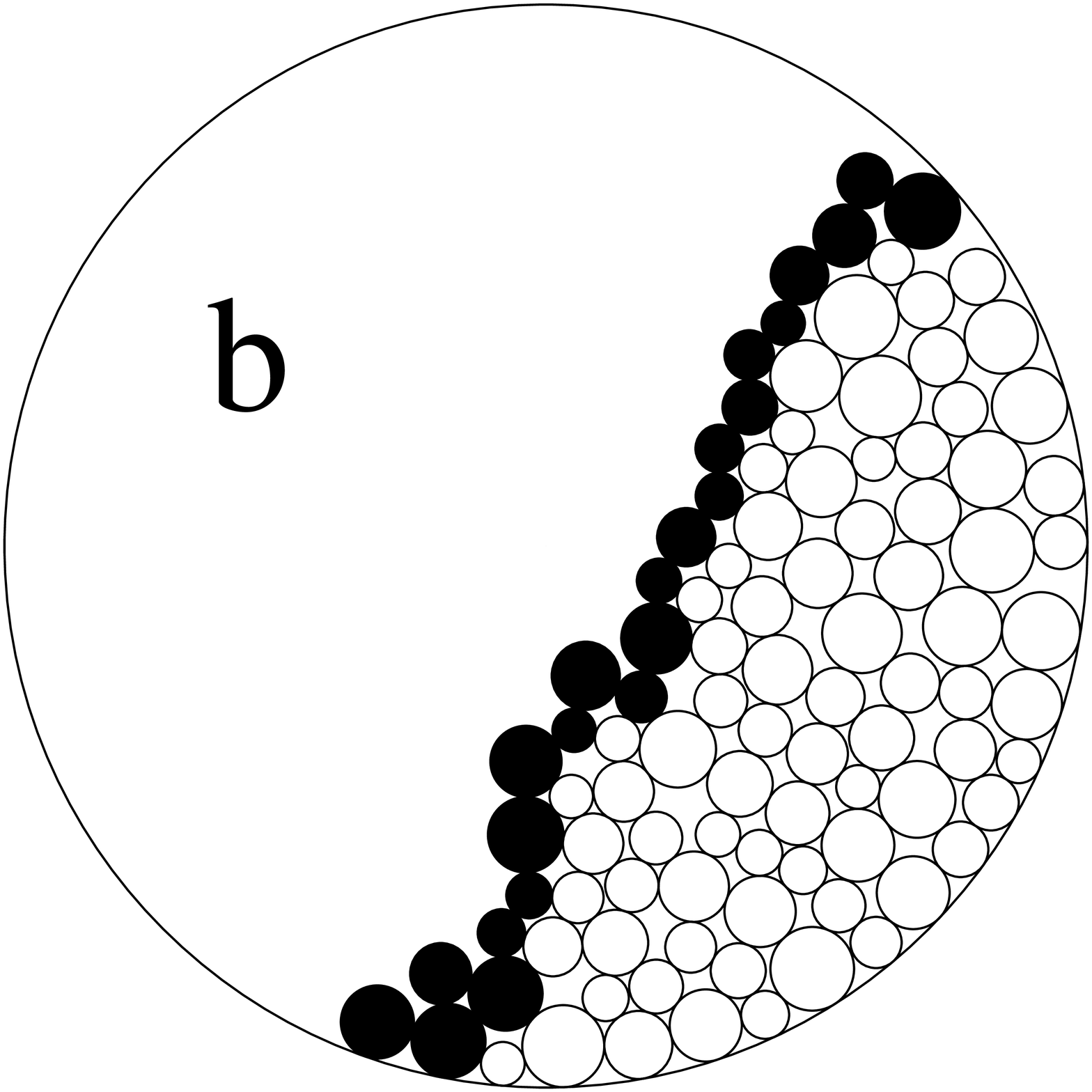}~
  \includegraphics[width=0.25\columnwidth]{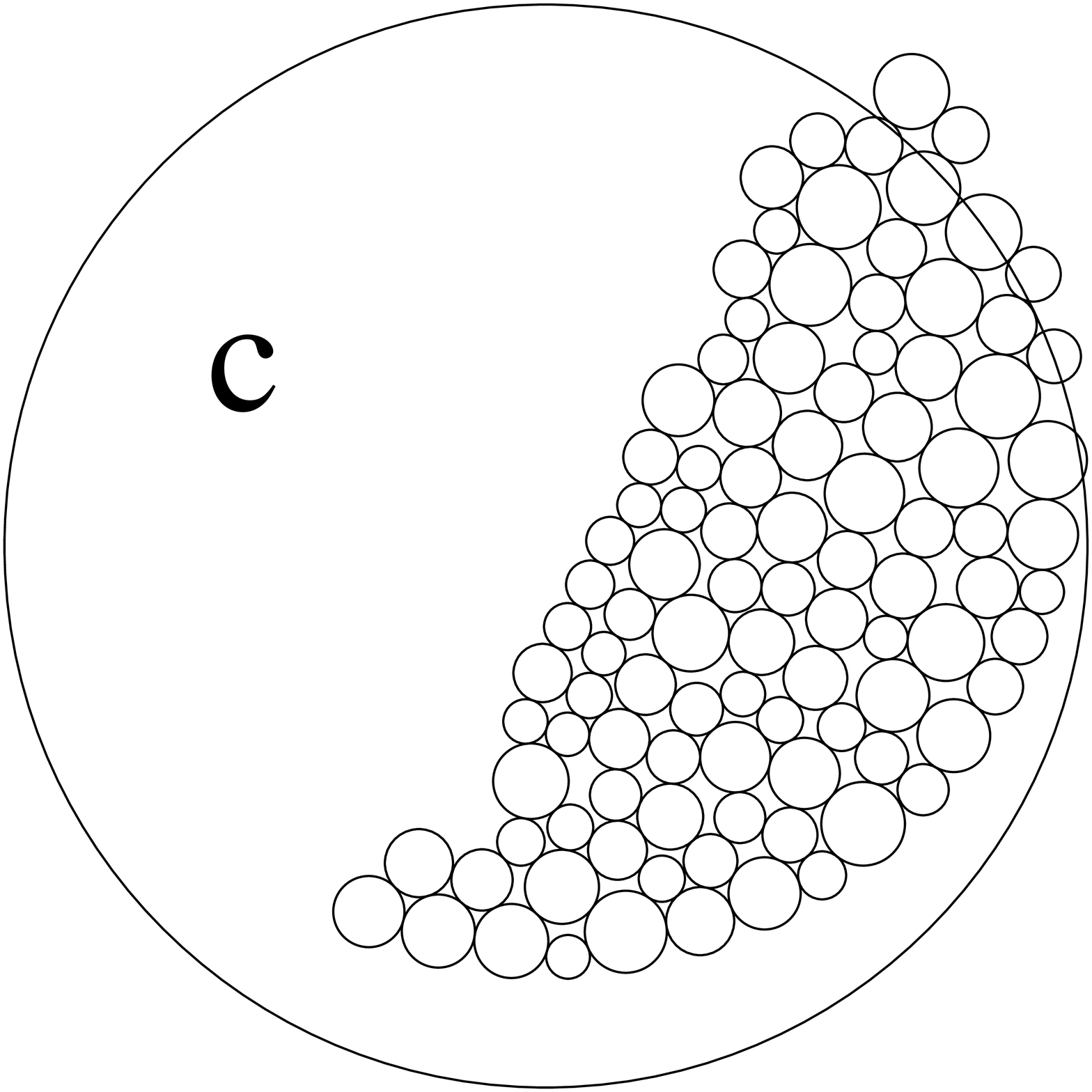}
  }
\vspace*{0.2cm}
\centerline{
  \includegraphics[width=0.25\columnwidth]{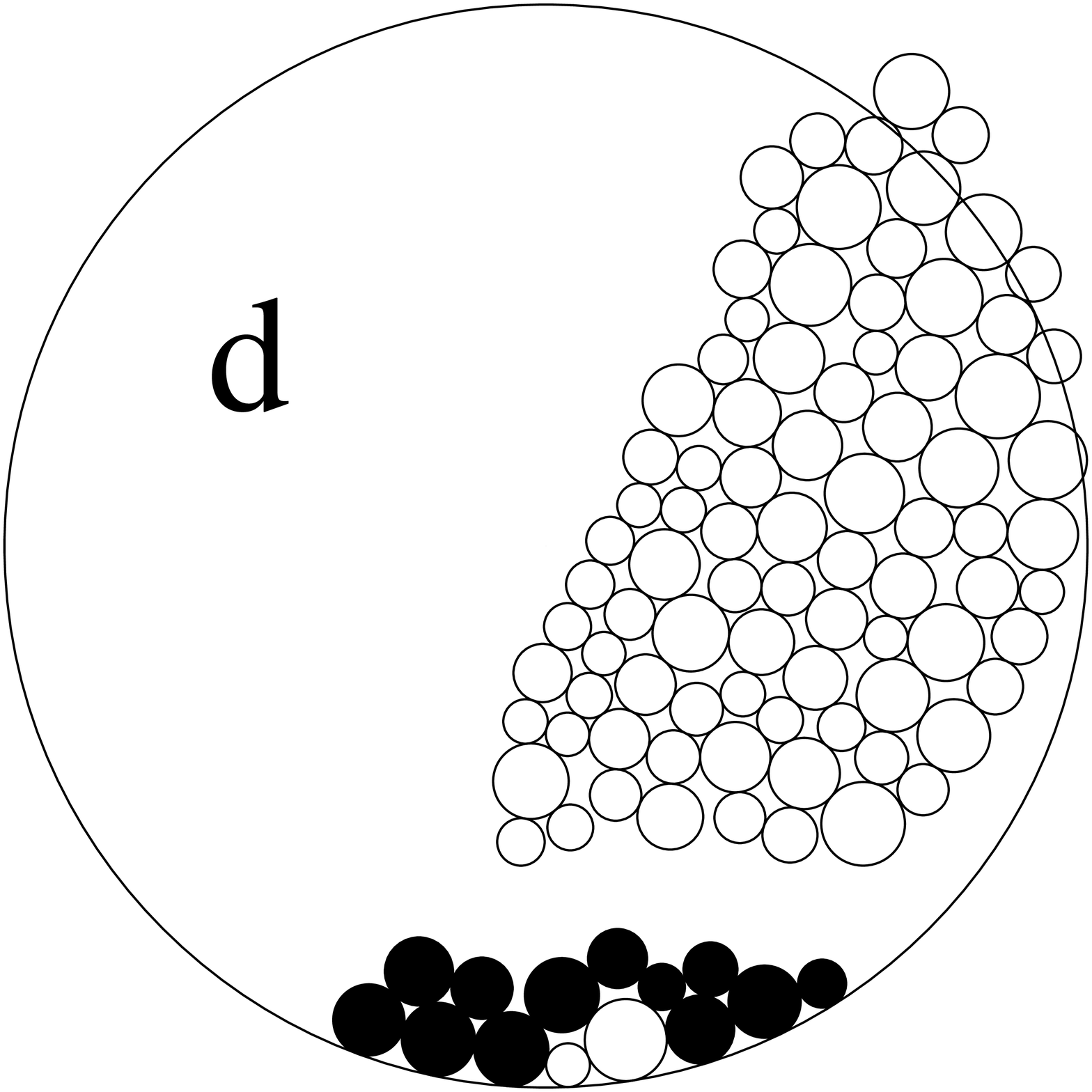}~
  \includegraphics[width=0.25\columnwidth]{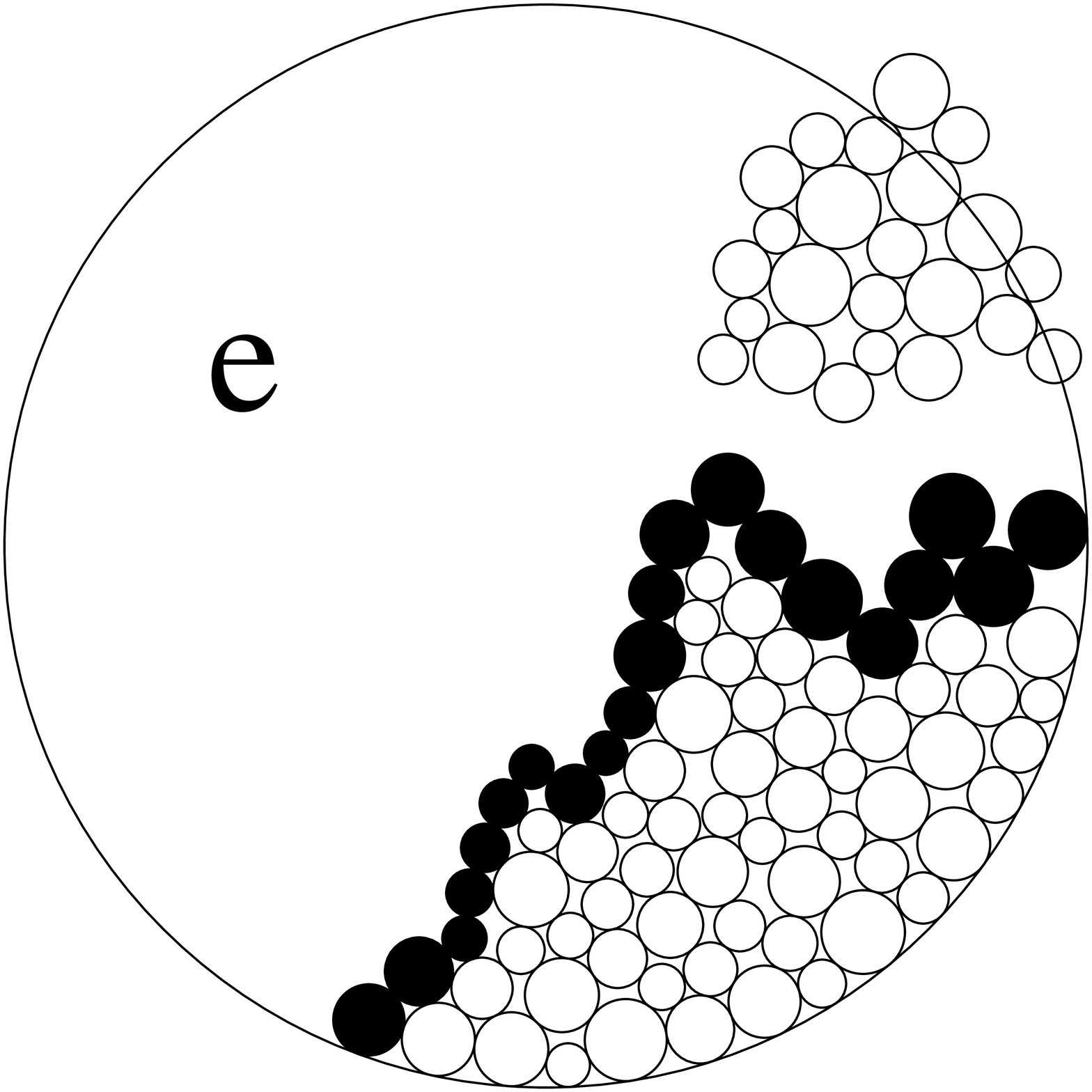}~
  \includegraphics[width=0.25\columnwidth]{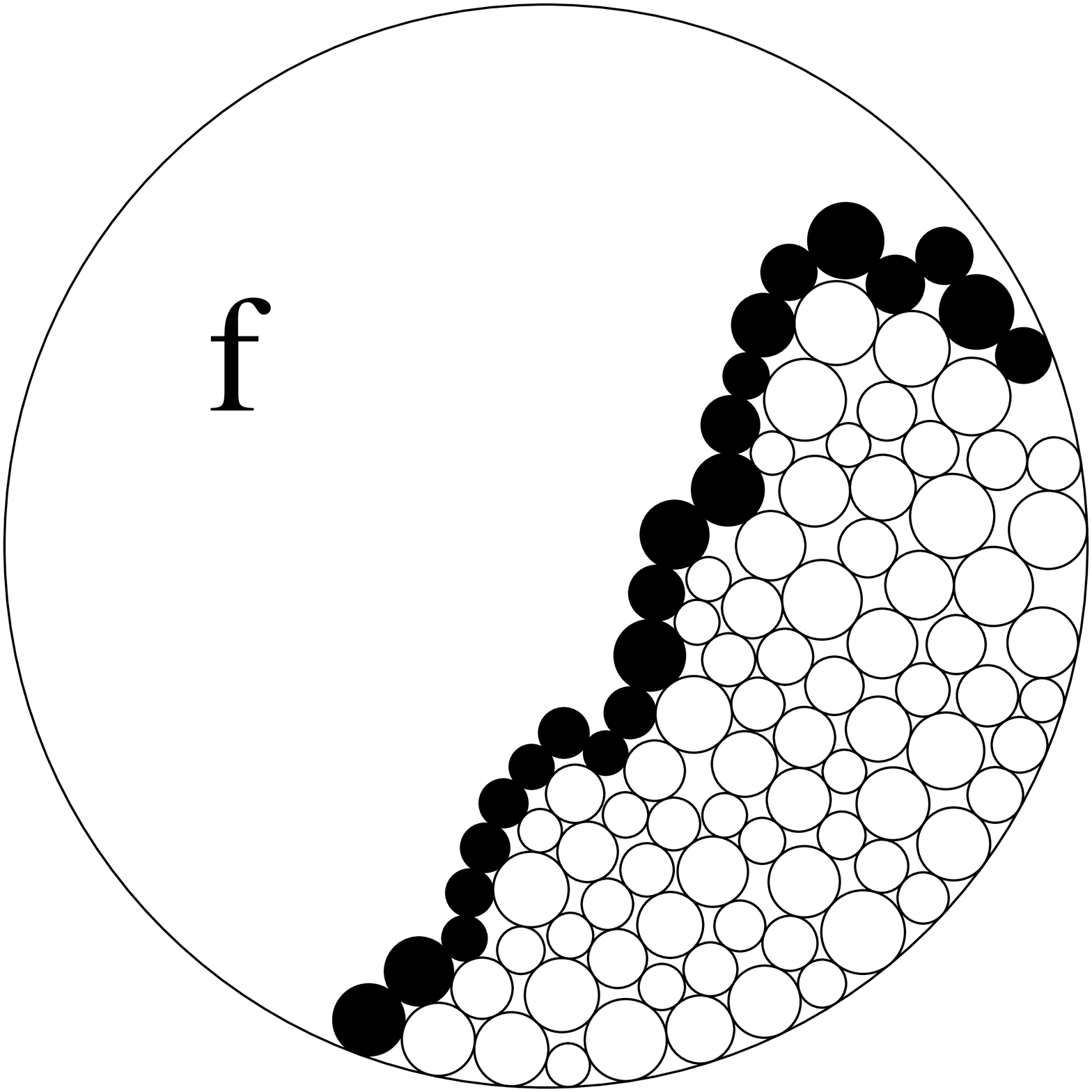}
  }
\caption{Dynamic BTR for the example of a slowly rotating cylinder. ({\em a})
  The system at time $t$. ({\em b}) The cylinder is rotated (angle of rotation
  appears exaggerated). ({\em c}) All particles are lifted by $R_{\rm
    cyl}/10$. ({\em d, e}) BTR, ({\em f}) Situation at time $t+\Delta t$ when
  all particles are deposited}
  \label{fig:VisscherAlgo}
\end{figure}

For the example of granular flow in a partially filled, slowly rotating
cylinder, the partition of the (continuous) dynamics is
\cite{baumann94,baumann95} (see Fig. \ref{fig:VisscherAlgo}):
\begin{enumerate}
\item {\em initialization:} Place the particles at random inside of the container.
\item {\em collective motion:} The positions of all particles follow the motion of the container for a small time step $\Delta t$. \label{loop}
\item {\em Preparation of the current time step:} Increase the $y$-coordinate of all particles by a constant, e.g., $R_{\rm cyl}/10$, with $R_{\rm cyl}$ being the cylinder radius. 
\item {\em BTR:} Apply BTR to the particles in sequence of increasing
  $y$-coordinate.
\item {\em Loop:} Increase the system time by $\Delta t$ and continue with step \ref{loop}.
\end{enumerate}

\begin{figure}[h!]
  \centering{
    \includegraphics[width=9cm,height=9cm]{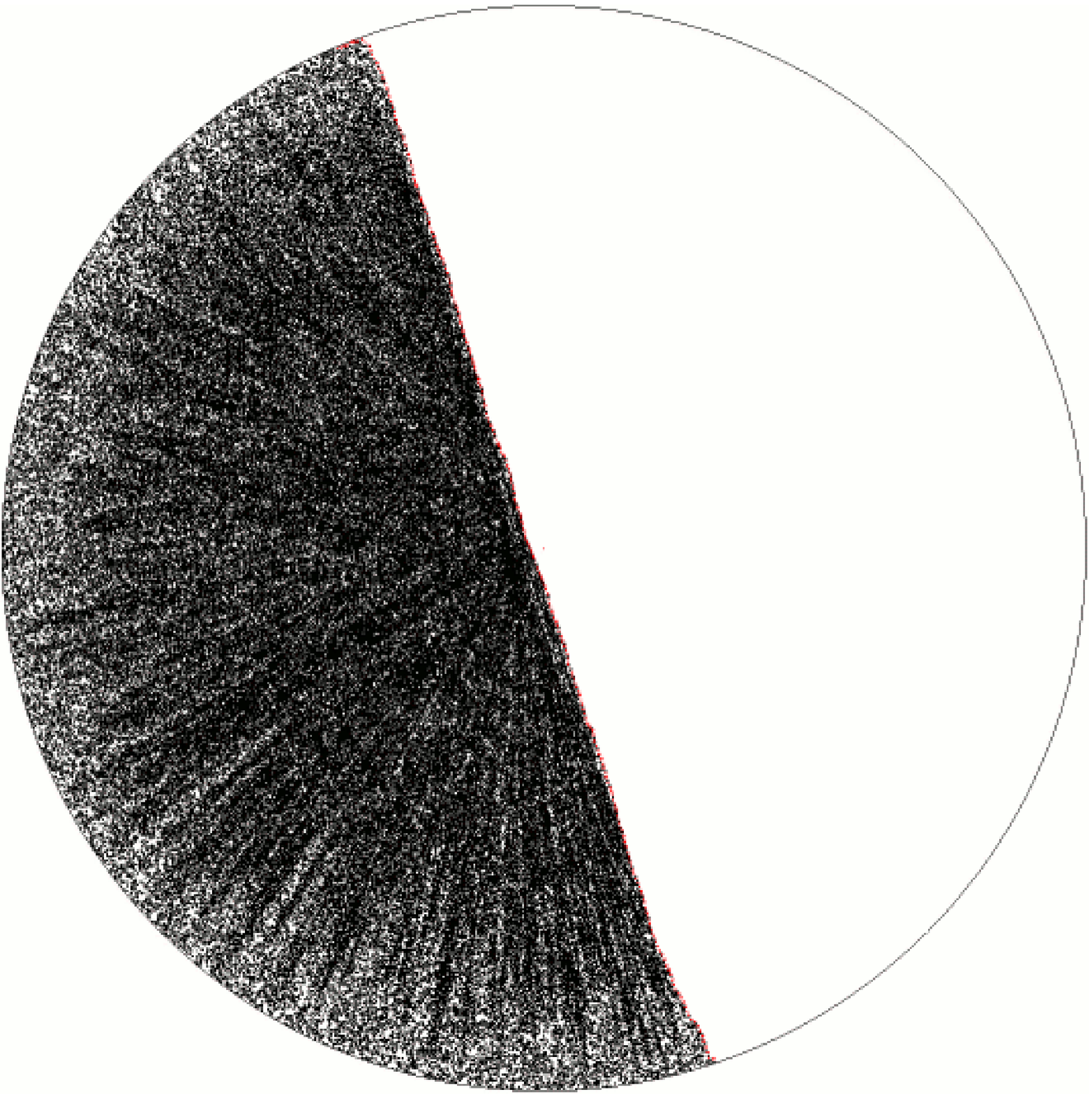}
    }
\vspace*{0.2cm}
  \centering{
    \includegraphics[width=4.5cm,height=4.5cm,bb=0 0 800 800,clip]{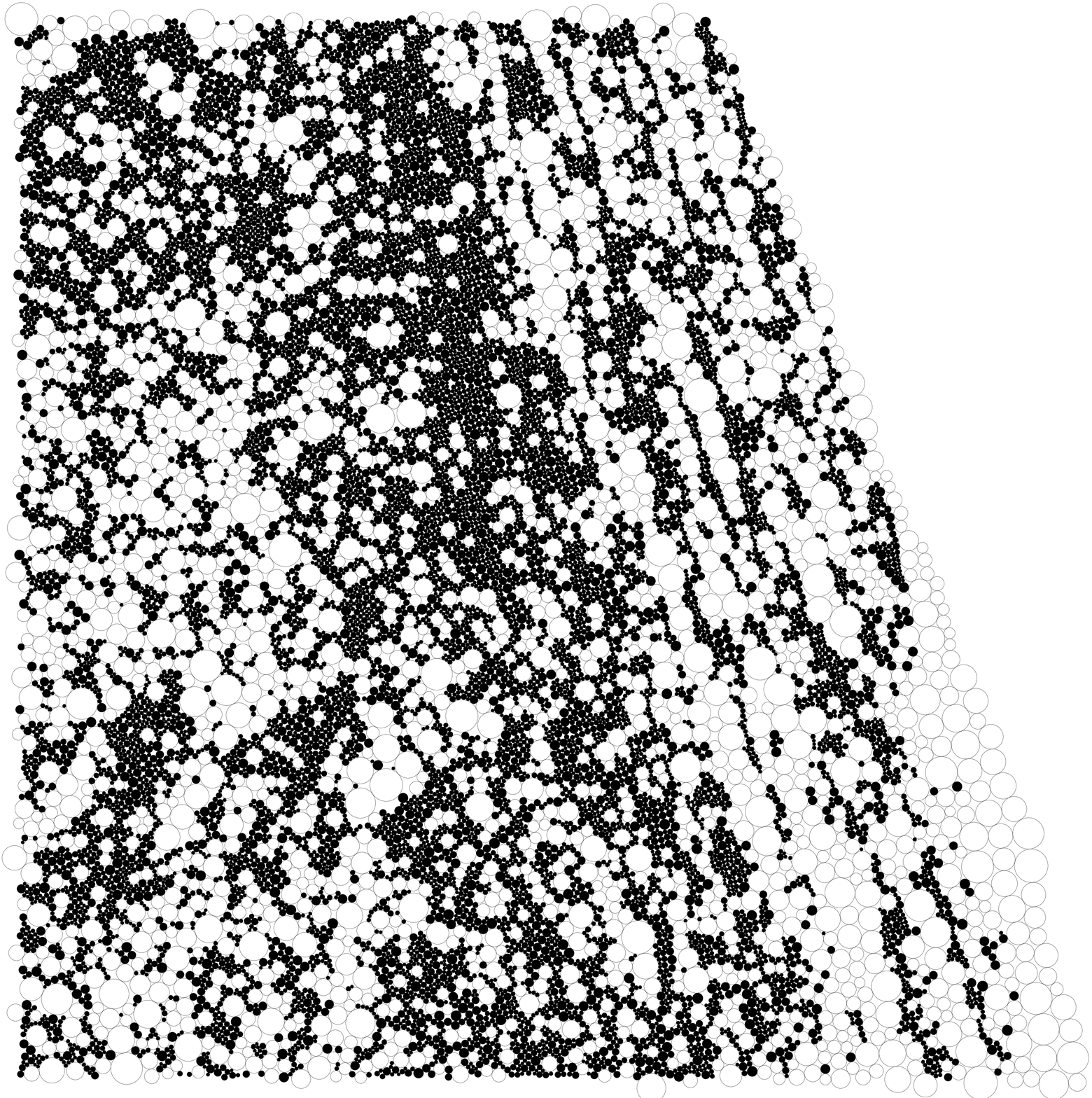}~
    \includegraphics[width=4.5cm,height=4.5cm,bb=0 0 800 800,clip]{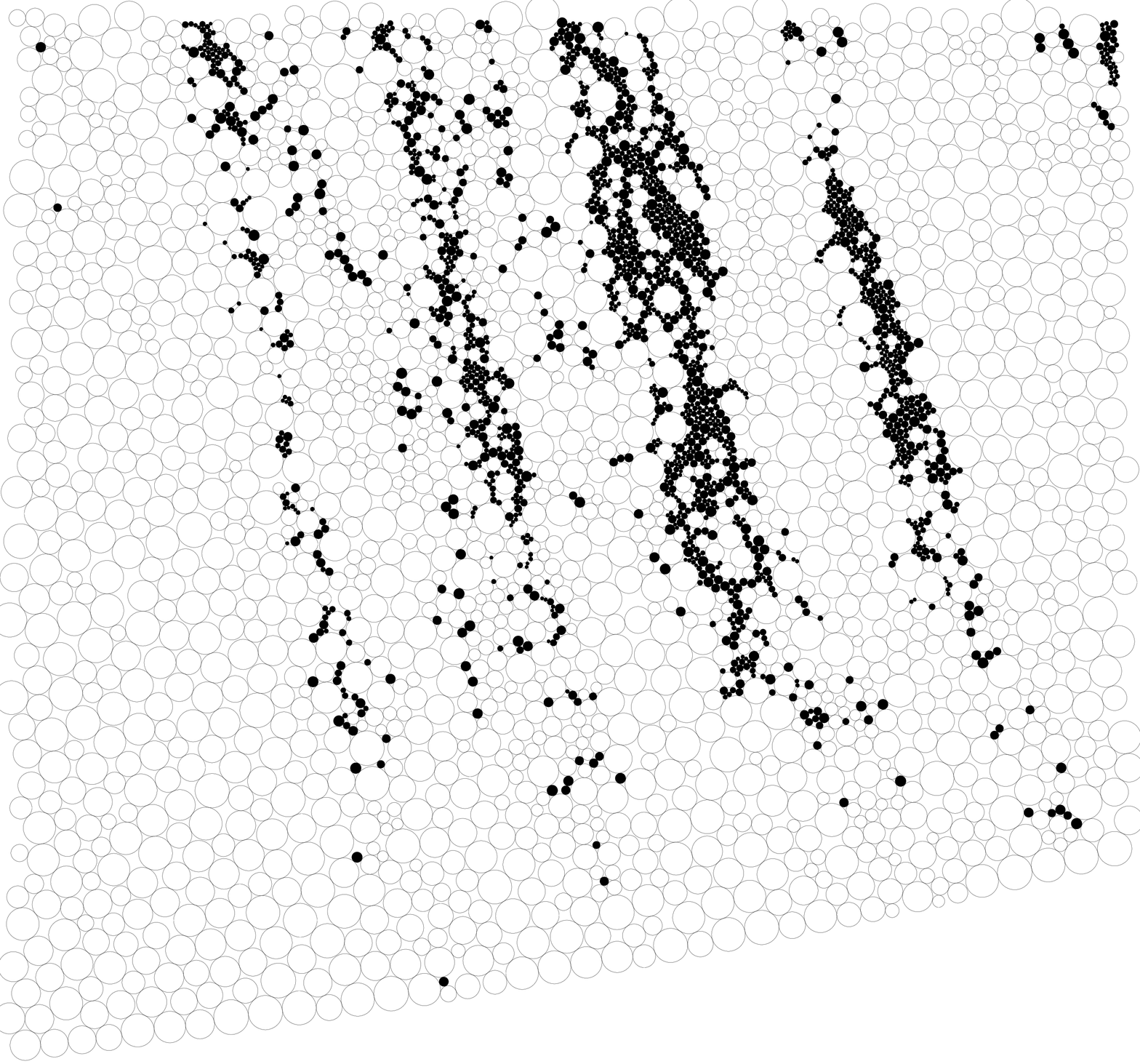}
    }  
  \caption{Snapshot of a slowly rotating cylinder filled with $N=10^6$ granular particles and close-ups. After one revolution consisting of 100 deposition steps, size segregation is clearly visible. For better presentation in the top figure only 30\% of the particles are drawn}
  \label{fig:Visscher.Zylinder}
\end{figure}

Figure \ref{fig:Visscher.Zylinder} 
shows a snapshot of a simulation of $N=10^6$ particles of different radii
$R_i\in(0.1,1)$\,cm in a cylinder of radius 70\,cm. The radii are chosen randomly in such a way that the 
the total mass of all particles from the interval $(R,R+{\rm d} R)$ is constant regardless of $R$.
We notice that the small particles are concentrated close to the center of the cylinder. This effect, which is observed also experimentally, was found in simulations in \cite{baumann93,baumann94}.

\section{Benchmarks of the Implementation and Critical Analysis of the Model}
\label{sec:bench}

BTR-simulations perform in general much faster than regular MD. Although BTR
was successfully applied to various large granular systems, this method is not
universal, and there are cases where physically incorrect behavior is observed \cite{barker93}. 

\begin{figure}[h!]
  \newcommand{\TSTPfigwidth}{2.9cm}
  \newcommand{\TSTPnegdist}{0.1cm}
  \centering
    \includegraphics[width=2.5cm,bb=46 70 457 512]{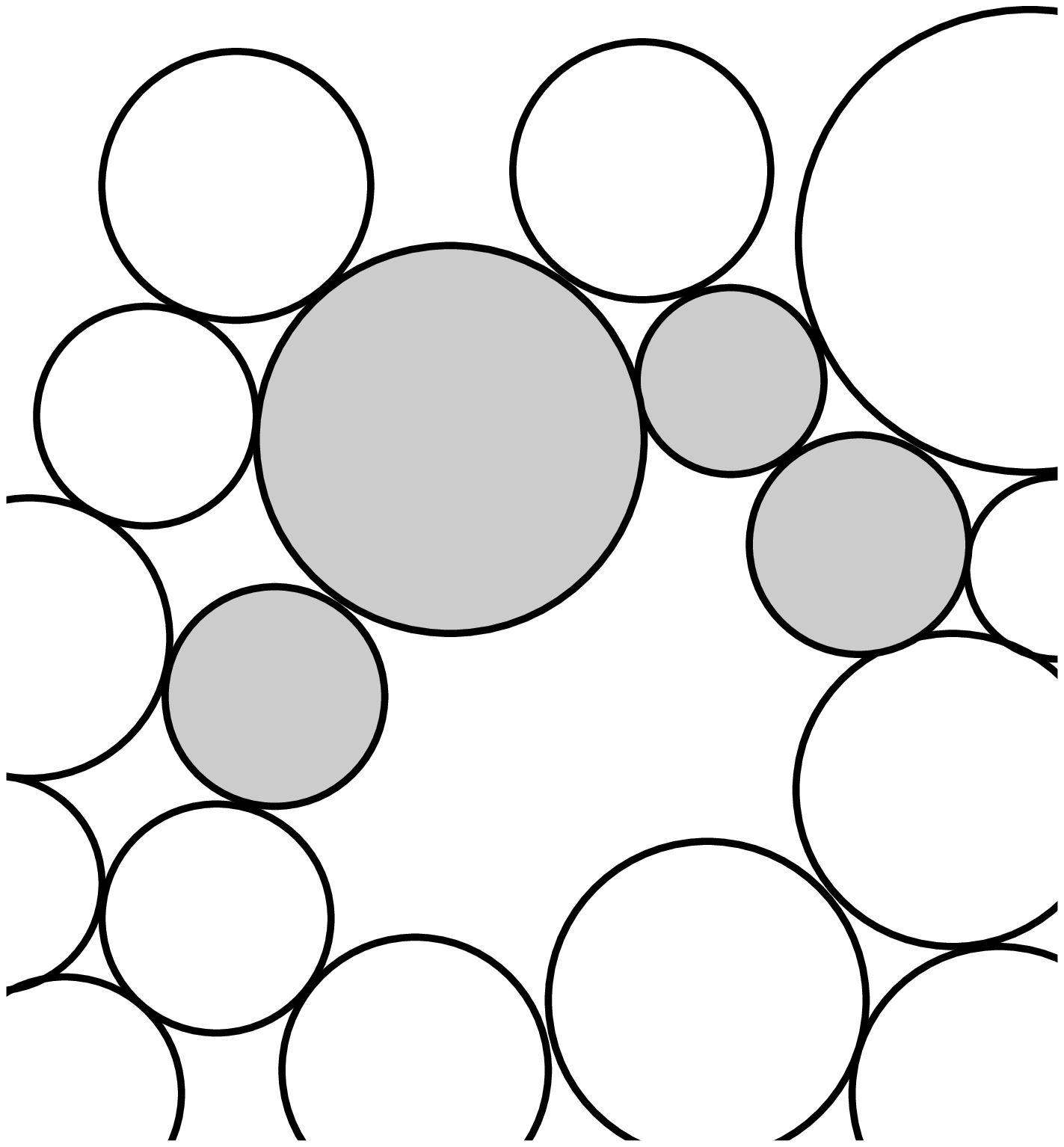}\hspace{\TSTPnegdist}
    \includegraphics[width=\TSTPfigwidth,bb=35 53 490 316]{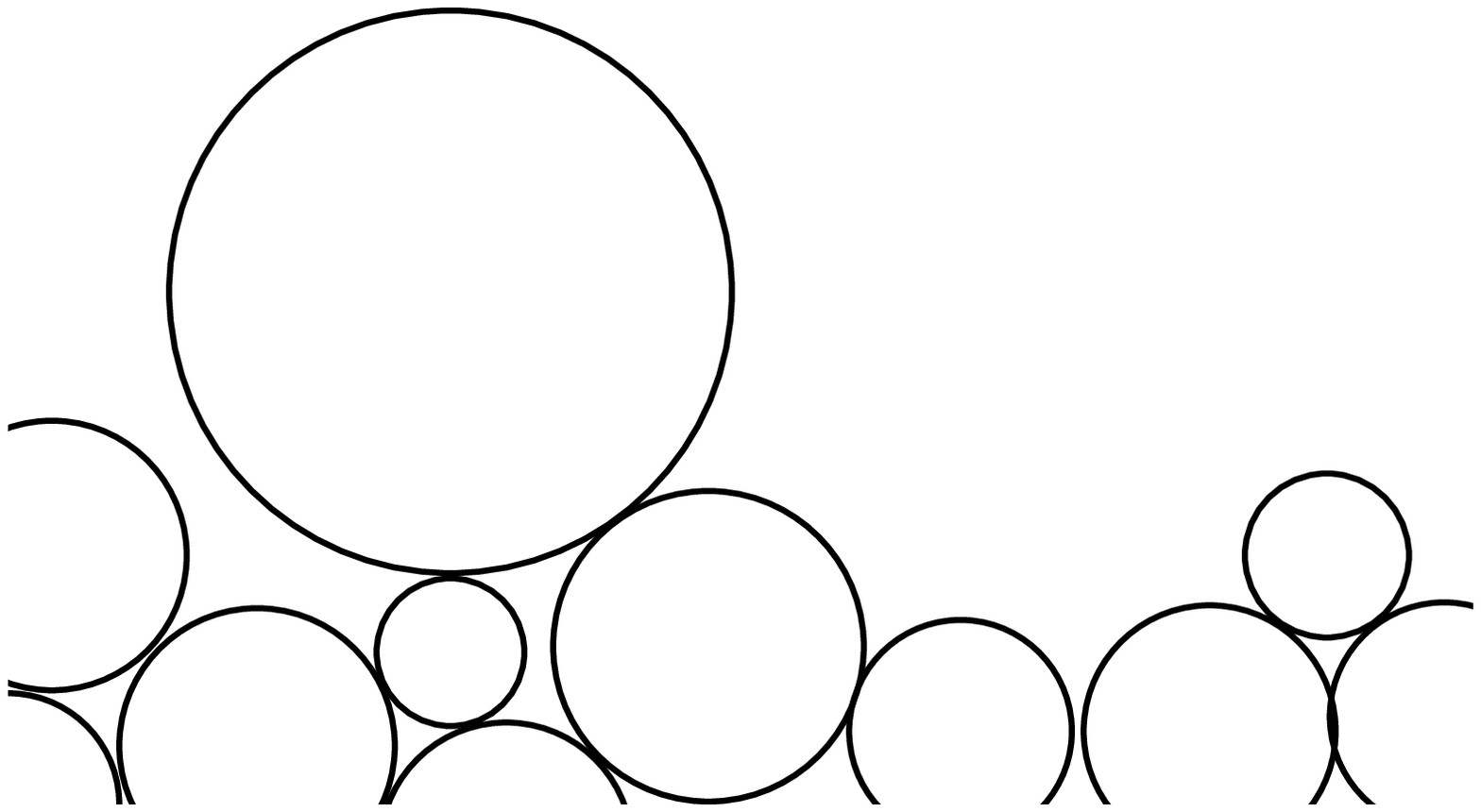}\hspace{\TSTPnegdist}
    \includegraphics[width=\TSTPfigwidth,bb=35 53 490 316]{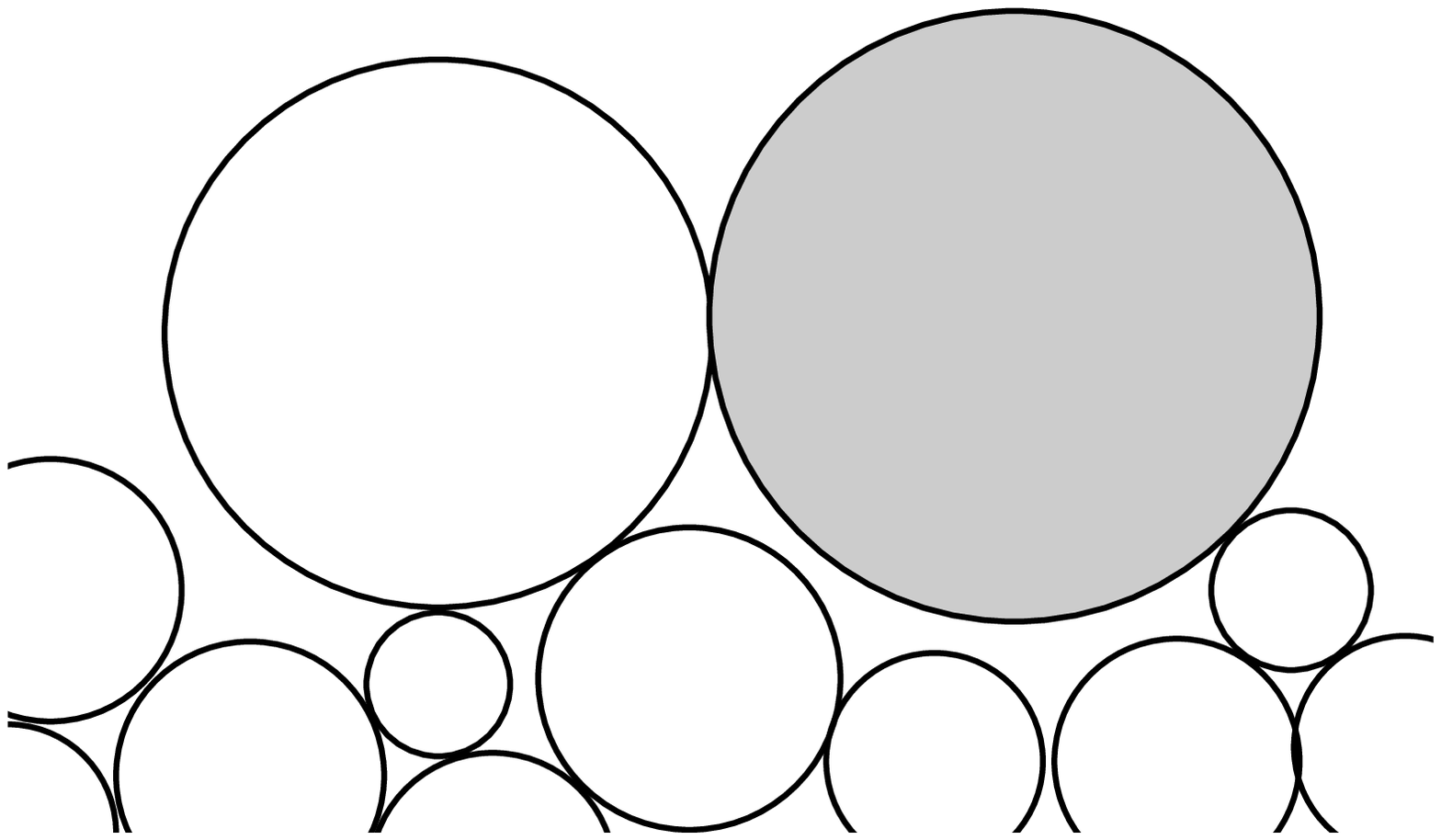}
     \caption{{\em left:} The configuration of particles is physically
  possible amd stable, but cannot be generated by BTR since none of the gray
  particles is in a stable position without the others, i.e., none of them
  could be deposited first. {\em middle and right:} BTR may generated this sequence, although it is not realistic since the configuration on the right-hand side is unstable}
  \label{fig:VBUnsinn}
\end{figure}

A deposited particle does not move under the influence of particles deposited
later, even if the particle suffers (in the realistic system) violent
collisions. The motion of the particles is, thus, not governed by Newton's
equation of motion. Instead, each single grain performs an overdamped motion
in a complicated potential landscape comprising the already deposited
grains. This is equivalent to disregarding inertial forces and moments. Even
from a very basic and intuitive concept of classical mechanics it is clear
that this algorithm cannot describe the granular many-body problem in a
general way. It should be regarded as a compromise between computing power
requirements and realistic description of physical reality. Figure
\ref{fig:VBUnsinn} demonstrates that BTR may lead to non-physical descriptions. 

\begin{figure}[htbp]
  \centerline{\includegraphics[width=0.8\columnwidth,clip]{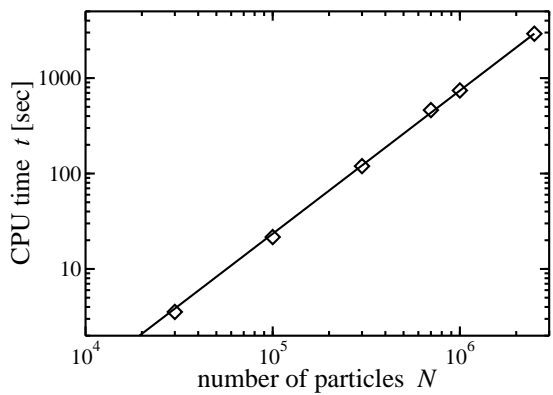}}
  \caption{CPU time to build a heap of $N$ particles using BTR. The data points (diamonds) are in close agreement with a powerlaw $t\propto N^{3/2}$ (solid line) as discussed in the text}
  \label{fig:bench}
\end{figure}

Nevertheless, {\em if} BTR is applicable, it leads to a great increase of
performance as compared with MD. Figure \ref{fig:bench} shows the CPU time to
deposit a heap if $N$ particles, using a desktop computer (Intel Pentium 4, 3GHz). The program
code is available at \cite{source}. 

The algorithm may yield satisfactory results when treating systems
where the boundary conditions change only very slowly. In such cases MD is frequently inefficient. In particular, for problems where the computation of the trajectories of individual particles is less important, the algorithm can be seen as a good compromise between efficiency and precision of the result.

\section*{Acknowledgments}
We thank Dietrich Wolf for discussion on the simulation of nano-powders.
This work was supported by German Science Foundation via grant Po472/10-1.

\bibliographystyle{abbrv}
\bibliography{SchwagerPoeschel}

\begin{thebibliography}{10}

\bibitem{source}
The source code of the described implementation of BTR can be obtained at
  http://bioinf.charite.de/cgd/.

\bibitem{Allen:1962}
J.~R. Allen.
\newblock Asymmetrical ripple marks and the origin of cross-stratification.
\newblock {\em Nature}, 194:167, 1962.

\bibitem{barker93}
G.~C. Barker, A.~Mehta, and M.~J. Grimson.
\newblock Comment on ``three-dimensional model for particle size segregation by
  shaking''.
\newblock {\em Phys. Rev. Lett.}, 70:2194, 1993.

\bibitem{baumann94}
G.~Baumann, I.~J\'anosi, and D.~E. Wolf.
\newblock Particle trajectories and segregation in a two-dimensional rotating
  drum.
\newblock {\em Europhys. Lett.}, 27:203, 1994.

\bibitem{baumann95}
G.~Baumann, I.~M. J\'anosi, and D.~E. Wolf.
\newblock Surface properties and flow of granular material in a 2d rotating
  drum model.
\newblock {\em Phys. Rev. E}, 51:1879, 1995.

\bibitem{baumann93}
G.~Baumann, E.~Jobs, and D.~E. Wolf.
\newblock Granular cocktail rotated and shaken.
\newblock {\em Fractals}, 1:767, 1993.

\bibitem{BehringerJenkins:1997}
R.~P. Behringer and J.~T. Jenkins, editors.
\newblock {\em Powders and Grains'97}, Rotterdam, 1997. Balkema.

\bibitem{BoutreuxDeGennes:1996}
T.~Boutreux and P.~G. de~Gennes.
\newblock Surface flows if granular mixtures: {I}. {G}eneral principles and
  minimal model.
\newblock {\em J. Physique I}, 6:1295, 1996.

\bibitem{BoutreuxDeGennes:1997}
T.~Boutreux and P.~G. de~Gennes.
\newblock Surface flow of granular mixtures.
\newblock In Behringer and Jenkins \cite{BehringerJenkins:1997}, page 439.

\bibitem{Brown:1939}
R.~L. Brown.
\newblock The fundamental principles of segregation.
\newblock {\em J. Inst. Fuel}, 13:15, 1939.

\bibitem{Goldhirsch:1999}
I.~Goldhirsch.
\newblock Granular gases: {P}robing the boundaries of hydrodynamics.
\newblock In T.~P\"oschel and S.~Luding, editors, {\em Granular Gases}, volume
  564 of {\em Lecture Notes in Physics}, page~79. Springer, Berlin, 2001.

\bibitem{GrasselliHerrmann:1997a}
Y.~Grasselli and H.~J. Herrmann.
\newblock Experimental study of granular stratification.
\newblock {\em Granular Matter}, 1:43, 1998.

\bibitem{GrayHutter:1997}
J.~M. N.~T. Gray and K.~Hutter.
\newblock Pattern formation in granular avalanches.
\newblock {\em Cont. Mech. and Thermodyn.}, 9:341, 1997.

\bibitem{GrayHutter:1998}
J.~M. N.~T. Gray and K.~Hutter.
\newblock Physik granularer {L}awinen.
\newblock {\em Physikalische Bl\"atter}, 54:37, 1998.

\bibitem{gray00:_shock}
J.~M. N.~T. Gray, Y.~C. Tai, and K.~Hutter.
\newblock Shock waves and particle size segregation in shallow granular flows.
\newblock In A.~D. Rosato and D.~L. Blackmore, editors, {\em IUTAM Symposium on
  Segregation in Granular Materials}, page 269, Dordrecht, 2000. Kluwer.

\bibitem{JulienLanRaslan:1997}
P.~Y. Julien, Y.~Q. Lan, and Y.~Raslan.
\newblock Experimental mechanics of sand stratification.
\newblock In Behringer and Jenkins \cite{BehringerJenkins:1997}, page 487.

\bibitem{jullien87b}
R.~Jullien and P.~Meakin.
\newblock Simple three-dimensional models for ballistic deposition with
  restructuring.
\newblock {\em Europhys. Lett.}, 4:1385, 1987.

\bibitem{jullien88}
R.~Jullien and P.~Meakin.
\newblock Ballistic deposition and segregation of polydisperse spheres.
\newblock {\em Europhys. Lett.}, 6:629, 1988.

\bibitem{jullien92}
R.~Jullien and P.~Meakin.
\newblock Three-dimensional model for particle-size segregation by shaking.
\newblock {\em Phys. Rev. Lett.}, 69:640, 1992.

\bibitem{jullien93b}
R.~Jullien, P.~Meakin, and A.~Pavlovitch.
\newblock Particle size segregation by shaking in two-dimensional disc
  packings.
\newblock {\em Europhys. Lett.}, 22:523, 1993.

\bibitem{KoeppeEnzKakalios:1997}
J.~Koeppe, M.~Enz, and J.~Kakalios.
\newblock Avalanche segregation of granular media.
\newblock In Behringer and Jenkins \cite{BehringerJenkins:1997}, page 443.

\bibitem{LitwiniszynCiTong:1963}
J.~Litwiniszyn and L.~Ci-Tong.
\newblock The phenomenon of segregation of grains of a loose medium when shaped
  in the form of a rotational half cone.
\newblock {\em Bull. de L'Acad\'emie Polonaise des Sciences, S\'erie des
  sciences techniques}, 11:169, 1963.

\bibitem{MakseCizeauStanley:1997}
H.~Makse, P.~Cizeau, and H.~E. Stanley.
\newblock Possible stratification mechanism in granular mixtures.
\newblock {\em Phys. Rev. Lett.}, 78:3298, 1997.

\bibitem{MakseCizeauStanley:1998}
H.~Makse, P.~Cizeau, and H.~E. Stanley.
\newblock Modeling stratification in two-dimensional sandpiles.
\newblock {\em Physica A}, 249:391, 1998.

\bibitem{Makse:1997}
H.~A. Makse.
\newblock Stratification instability in granular flows.
\newblock {\em Phys. Rev. E}, 56:7008, 1997.

\bibitem{MakseEtAl:1996}
H.~A. Makse, S.~Havlin, P.~C. Ivanov, P.~R. King, S.~Prakash, and H.~E.
  Stanley.
\newblock Pattern formation in sedimentary rocks: {C}onnectivity, permeability,
  and spatial correlations.
\newblock {\em Physica A}, 233:587, 1996.

\bibitem{MakseHavlinKingStanley:1997}
H.~A. Makse, S.~Havlin, P.~R. King, and H.~E. Stanley.
\newblock Novel pattern formation in granular matter.
\newblock In L.~Schimansky-Geier and T.~P\"oschel, editors, {\em Stochastic
  Dynamics}, Lecture Notes in Physics, page 319, Berlin Heidelberg New York,
  1997. Springer.

\bibitem{MakseHavlinKingStanley:1997a}
H.~A. Makse, S.~Havlin, P.~R. King, and H.~E. Stanley.
\newblock Spontaneous stratification in granular mixtures.
\newblock {\em Nature}, 386:379, 1997.

\bibitem{MakseHerrmann:1997}
H.~A. Makse and H.~J. Herrmann.
\newblock Microscopic model for granular stratification and segregation.
\newblock {\em Europhys. Lett.}, 43:1, 1998.

\bibitem{Meakin:1990}
P.~Meakin.
\newblock A simple two-dimensional model for particle segregation.
\newblock {\em Physica A}, 163:733, 1990.

\bibitem{ORourke}
J.~O'Rourke.
\newblock {\em Computational Geometry}.
\newblock Cambridge University Class, 2000.

\bibitem{algo}
T.~P\"oschel and T.~Schwager.
\newblock {\em Computational Granular Dynamics: Models and Algorithms}.
\newblock Springer, Berlin, Heidelberg, New-York, 2005.

\bibitem{PressVetterlingTeukolskyFlannery:1988}
W.~H. Press, W.~T. Vetterling, S.~A. Teukolsky, and B.~P. Flannery.
\newblock {\em Numerical Recipes}.
\newblock Cambridge University Press, Cambridge, 1988.

\bibitem{SchwagerPoeschel:inprep}
T.~Schwager, D.~Wolf, and T.~P{\"o}schel.
\newblock in preparation.

\bibitem{Sorby:1859}
H.~C. Sorby.
\newblock On the structures produced by the currents present during the
  deposition of stratified rocks.
\newblock {\em The Geologist}, 2:137, 1859.

\bibitem{VisscherBolsterli:1972}
W.~M. Visscher and M.~Bolsterli.
\newblock Random packing of equal and unequal spheres in two and three
  dimensions.
\newblock {\em Nature}, 239:504, 1972.

\bibitem{Williams:1963}
J.~C. Williams.
\newblock The segregation of powders and granular materials.
\newblock {\em Univ. Sheffield Fuel Soc. J.}, 14:29, 1963.

\bibitem{Williams:1976}
J.~C. Williams.
\newblock The segregation of particulate materials. {A} review.
\newblock {\em Powder Techn.}, 15:245, 1976.

\end{thebibliography}

\end{document}